\documentstyle[12pt,epsfig]{article}

\newcommand{\tr}{\mathop{\mathrm{tr}}\nolimits}
\newcommand{\diag}{\mathop{\mathrm{diag}}\nolimits}
\newcommand{\agt}{\,\rlap{\lower 3.5 pt \hbox{$\mathchar \sim$}} \raise 1pt
 \hbox {$>$}\,}
\newcommand{\alt}{\,\rlap{\lower 3.5 pt \hbox{$\mathchar \sim$}} \raise 1pt
 \hbox {$<$}\,}

\catcode`@=11
\newcount\@tempcntc
\def\@citex[#1]#2{\if@filesw\immediate\write\@auxout{\string\citation{#2}}\fi
  \@tempcnta\z@\@tempcntb\m@ne\def\@citea{}\@cite{\@for\@citeb:=#2\do
    {\@ifundefined
       {b@\@citeb}{\@citeo\@tempcntb\m@ne\@citea\def\@citea{,}{\bf ?}\@warning
       {Citation `\@citeb' on page \thepage \space undefined}}%
    {\setbox\z@\hbox{\global\@tempcntc0\csname b@\@citeb\endcsname\relax}%
     \ifnum\@tempcntc=\z@ \@citeo\@tempcntb\m@ne
       \@citea\def\@citea{,}\hbox{\csname b@\@citeb\endcsname}%
     \else
      \advance\@tempcntb\@ne
      \ifnum\@tempcntb=\@tempcntc
      \else\advance\@tempcntb\m@ne\@citeo
      \@tempcnta\@tempcntc\@tempcntb\@tempcntc\fi\fi}}\@citeo}{#1}}
\def\@citeo{\ifnum\@tempcnta>\@tempcntb\else\@citea\def\@citea{,}%
  \ifnum\@tempcnta=\@tempcntb\the\@tempcnta\else
   {\advance\@tempcnta\@ne\ifnum\@tempcnta=\@tempcntb \else \def\@citea{--}\fi
    \advance\@tempcnta\m@ne\the\@tempcnta\@citea\the\@tempcntb}\fi\fi}
\catcode`@=12
\begin{document}

\title{\vskip-3cm{\baselineskip14pt
\centerline{\normalsize DESY 01-196\hfill ISSN 0418-9833}
\centerline{\normalsize MPI/PhT/2001-29\hfill}
\centerline{\normalsize hep-ph/0112199\hfill}
\centerline{\normalsize August 2001\hfill}
}
\vskip1.5cm
$J/\psi$ Inclusive Production in $ep$ Deep-Inelastic Scattering at DESY HERA}
\author{
{\sc Bernd A. Kniehl}\thanks{Permanent address: II. Institut f\"ur
Theoretische Physik, Universit\"at Hamburg, Luruper Chaussee 149, 22761
Hamburg, Germany.}
\\
{\normalsize Max-Planck-Institut f\"ur Physik (Werner-Heisenberg-Institut),}\\
{\normalsize F\"ohringer Ring 6, 80805 Munich, Germany}\\
\\
{\sc Lennart Zwirner}\\
{\normalsize II. Institut f\"ur Theoretische Physik, Universit\"at Hamburg,}\\
{\normalsize Luruper Chaussee 149, 22761 Hamburg, Germany}}

\date{}

\maketitle

\thispagestyle{empty}

\begin{abstract}
We calculate the cross section of $J/\psi$ plus jet associated production in
$ep$ deep-inelastic scattering within the factorization formalism of
nonrelativistic quantum chromodynamics.
Our analytic results disagree with previous analyses, both for the 
colour-singlet and colour-octet channels.
Our theoretical predictions agree reasonably well with recent data taken by
the H1 Collaboration at DESY HERA, significantly better than those obtained
within the colour-singlet model.

\medskip

\end{abstract}

\newpage

\section{Introduction}

Since its discovery in 1974, the $J/\psi$ meson has provided a useful
laboratory for quantitative tests of quantum chromodynamics (QCD) and, in
particular, of the interplay of perturbative and nonperturbative phenomena.
The factorization formalism of nonrelativistic QCD (NRQCD) \cite{bbl} provides
a rigorous theoretical framework for the description of heavy-quarkonium
production and decay.
This formalism implies a separation of short-distance coefficients, which can 
be calculated perturbatively as expansions in the strong-coupling constant
$\alpha_s$, from long-distance matrix elements (MEs), which must be extracted
from experiment.
The relative importance of the latter can be estimated by means of velocity
scaling rules, i.e.\ the MEs are predicted to scale with a definite power of
the heavy-quark ($Q$) velocity $v$ in the limit $v\ll1$.
In this way, the theoretical predictions are organized as double expansions in
$\alpha_s$ and $v$.
A crucial feature of this formalism is that it takes into account the complete
structure of the $Q\overline{Q}$ Fock space, which is spanned by the states
$n={}^{2S+1}L_J^{(c)}$ with definite spin $S$, orbital angular momentum $L$,
total angular momentum $J$, and colour multiplicity $c=1,8$.
In particular, this formalism predicts the existence of colour-octet (CO)
processes in nature.
This means that $Q\overline{Q}$ pairs are produced at short distances in
CO states and subsequently evolve into physical, colour-singlet (CS) quarkonia
by the nonperturbative emission of soft gluons.
In the limit $v\to 0$, the traditional CS model (CSM) \cite{ber,bai} is
recovered.
The greatest triumph of this formalism was that it was able to correctly 
describe \cite{ebr,cho} the cross section of inclusive charmonium
hadroproduction measured in $p\overline{p}$ collisions at the Fermilab
Tevatron \cite{abe}, which had turned out to be more than one order of
magnitude in excess of the theoretical prediction based on the CSM.

In order to convincingly establish the phenomenological significance of the
CO processes, it is indispensable to identify them in other kinds of
high-energy experiments as well.
Studies of charmonium production in $ep$ photoproduction, $ep$ and $\nu N$
deep-inelastic scattering (DIS), $e^+e^-$ annihilation, $\gamma\gamma$
collisions, and $b$-hadron decays may be found in the literature; see
Ref.~\cite{bra} and references cited therein.
Furthermore, the polarization of charmonium, which also provides a sensitive
probe of CO processes, was investigated \cite{ben,bkv,bkl}.
None of these studies was able to prove or disprove the NRQCD factorization
hypothesis.

In this paper, we revisit $J/\psi$ inclusive production in $ep$ DIS.
In order to avoid kinematic overlap with diffractive production, which cannot
yet be reliably described within purely perturbative QCD, we require that the
$J/\psi$ meson be produced in association with a hadron jet $j$, i.e.\ we
consider the process $e+p\to e+J/\psi+j+X$, where $X$ denotes the proton
remnant.
In this way, the inelasticity variable $z$, which measures the fraction of the 
virtual-photon ($\gamma^\star$) energy transferred to the $J/\psi$ meson in
the proton rest frame, can take values below unity, away from the endpoint
$z=1$, where diffractive production takes place.
At the same time, the $J/\psi$ meson can acquire finite transverse momentum
$p_T^\star$ in the $\gamma^\star p$ centre-of-mass (CM) frame, and the
hadronic system $X^\prime$ consisting of the jet $j$ and the proton remnant
$X$ can acquire finite mass $M_{X^\prime}$.
By the same token, diffractive events can be eliminated from the experimental
data sample by applying appropriate acceptance cuts on $z$, $p_T^\star$, or
$M_{X^\prime}$.
Another possibility to suppress the diffractive background at $z\alt1$ would 
be to require that the photon virtuality $Q^2$ be sufficiently large
\cite{bro}.
However, then also the bulk of the nondiffractive signal would be sacrificed.

\begin{figure}[ht]
\begin{center}
\centerline{
\epsfig{figure=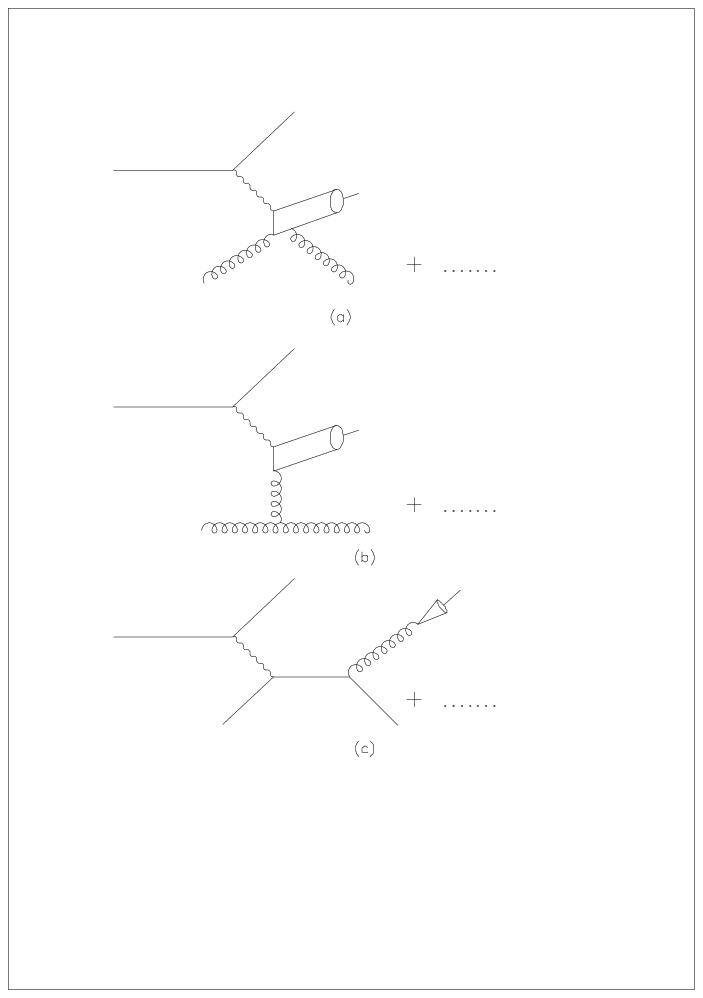,height=18cm,bbllx=214pt,bblly=121pt,bburx=317pt,%
bbury=311pt,clip=}
}
\caption{Representative Feynman diagrams for the partonic subprocesses
$e+a\to e+c\overline{c}[n]+a$, where $a=g,q,\overline{q}$ and
$n={}^3\!S_1^{(1)},{}^1\!S_0^{(8)},{}^3\!S_1^{(8)},{}^3\!P_J^{(8)}$.
There are six diagrams of the type shown in part (a), two ones of the type
shown in part (b), and two ones of the type shown in part (c).
There are two more diagrams that are obtained from the diagrams of the type
shown in part (b) by replacing the external gluon lines with quark ones.
The CS process only proceeds through the diagrams shown in part (a).}
\label{fig:fey}
\end{center}
\end{figure}

The leading CS ME of the $J/\psi$ meson is
$\left\langle{\cal O}^\psi\left[{}^3\!S_1^{(1)}\right]\right\rangle$, its
leading CO ones are\break
$\left\langle{\cal O}^\psi\left[{}^1\!S_0^{(8)}\right]\right\rangle$,
$\left\langle{\cal O}^\psi\left[{}^3\!S_1^{(8)}\right]\right\rangle$, and
$\left\langle{\cal O}^\psi\left[{}^3\!P_J^{(8)}\right]\right\rangle$, with
$J=0,1,2$.
At LO, we are thus led to consider the partonic subprocesses
$e+a\to e+c\overline{c}[n]+a$, where $a=g,q,\overline{q}$ and
$n={}^3\!S_1^{(1)},{}^1\!S_0^{(8)},{}^3\!S_1^{(8)},{}^3\!P_J^{(8)}$.
Here, $q$ runs over the light-quark flavours $u$, $d$, and $s$.
Notice that
$e+q(\overline{q})\to e+c\overline{c}\left[{}^3\!S_1^{(1)}\right]
+q(\overline{q})$ is forbidden because the charm-quark line is connected to
one gluon, which transfers colour to the $c\overline{c}$ pair.
Representative Feynman diagrams are depicted in Fig.~\ref{fig:fey}.
The corresponding cross sections are of ${\cal O}(\alpha^2\alpha_s^2)$.

There are several motivations for our study of $e+p\to e+J/\psi+j+X$ in DIS.
For one thing, the H1 Collaboration measured various distributions of this
process at DESY HERA \cite{mey,h1,mor}, which wait to be confronted with
appropriate theoretical predictions.
This allows for a particularly clean test of the NRQCD factorization
hypothesis, since the large photon virtuality $Q^2$ ensures that perturbative
QCD is applicable and that the resolved-photon contribution, which suffers
from our imperfect knowledge of the parton density functions (PDFs) of the
photon, is greatly suppressed.
Furthermore, at least within the CSM, this process provides a good handle on
the gluon PDF of the proton, which is less precisely known than the quark
ones.
In NRQCD, however, the extraction of the gluon PDF is somewhat aggravated by
the presence of the CO channels with a quark or antiquark in the initial state
and by the uncertainties associated with the CO MEs.
Fortunately, detailed inspection reveals that the relative importance of the
quark- and antiquark-induced channels is greatly damped in the HERA regime.
We return to this point at the end of Section~\ref{sec:three}.

On the other hand, in the case of $J/\psi$ inelastic photoproduction, with
$Q^2\approx0$, NRQCD with CO MEs tuned \cite{cho} to fit the Tevatron data
\cite{abe} predicts \cite{cac,ko} at leading order (LO) a distinct rise in
cross section as $z\to1$, which is not observed by the H1 \cite{aid} and ZEUS
\cite{bre} Collaborations at HERA.
This CO charmonium anomaly has cast doubts on the validity of the NRQCD
factorization hypothesis, which seems so indispensible to interpret the
Tevatron data in a meaningful way.
Although there are several interesting and promising ideas how to reconcile
this data with the NRQCD prediction, e.g.\ by including dominant higher-order
effects \cite{ano}, by introducing nonperturbative shape functions that resum
higher-order corrections related to the kinematics of soft-gluon radiation and
to the difference between the partonic and hadronic phase spaces \cite{rot},
or by endowing the partons inside the proton with intrinsic transverse
momentum ($k_T$) \cite{sri}, it is of great interest to find out if this
anomaly persists, at LO and without resorting to shape functions or $k_T$
effects, if $Q^2$ is increased to large values.

On the theoretical side, the cross sections of the partonic subprocesses under
consideration here constitute an essential ingredient for the calculation of
the next-to-leading-order (NLO) corrections to the inclusive cross section of
the reaction $e+p\to e+J/\psi+X$ in DIS, which, at LO, proceeds through the
partonic subprocesses $e+g\to e+c\overline{c}[n]$ with
$n={}^1\!S_0^{(8)},{}^3\!P_J^{(8)}$.
The cross sections of the latter are of ${\cal O}(\alpha^2\alpha_s)$ and may
be found in Eq.~(3) of Ref.~\cite{fle}.
In fact, integrating over the phase space of the massless final-state parton 
$a$, one obtains the real radiative corrections.
If $a=g$ and $n={}^1\!S_0^{(8)},{}^3\!P_J^{(8)}$, then they suffer from both
infrared (IR) singularities and collinear ones associated with the incoming
gluon.
The latter are factorized, at some factorization scale $M$, and absorbed into
the bare gluon PDF of the proton, so as to render it renormalized.
The IR singularities cancel when the real radiative corrections are combined
with the virtual ones.
Finally, the ultraviolet (UV) radiative corrections contained in the latter
are removed by renormalizing the couplings, masses, wave-functions, and
non-perturbative MEs appearing in the LO cross section of $e+p\to e+J/\psi+X$.

Finally, the literature contains mutually inconsistent formulas for the cross
section of the CS partonic subprocess
$e+g\to e+c\overline{c}\left[{}^3\!S_1^{(1)}\right]+g$
\cite{koe,gui,mer,kru,yua}.
On the other hand, there is only one paper specifying analytic results for the
cross sections of the CO partonic subprocesses enumerated above \cite{yua}, so
that an independent check seems to be in order.
We anticipate that we disagree with all published CS and CO formulas, except 
with the one referring to
$e+g\to e+c\overline{c}\left[{}^1\!S_0^{(8)}\right]+g$ \cite{yua}.

This paper is organized as follows.
In Section~\ref{sec:two}, we present, in analytic form, the cross sections of
the partonic subprocesses $e+a\to e+c\overline{c}[n]+a$ enumerated above to LO
in NRQCD and explain how to calculate from them the total cross section of
$e+p\to e+J/\psi+j+X$ in DIS and several distributions of phenomenological
interest.
Lengthy expressions are relegated to the Appendix.
In Section~\ref{sec:three}, we present our numerical results and compare them
with recent H1 data \cite{mey}.
Our conclusions are summarized in Section~\ref{sec:four}.

\section{Analytic results}
\label{sec:two}

\begin{figure}[ht]
\begin{center}
\centerline{
\epsfig{figure=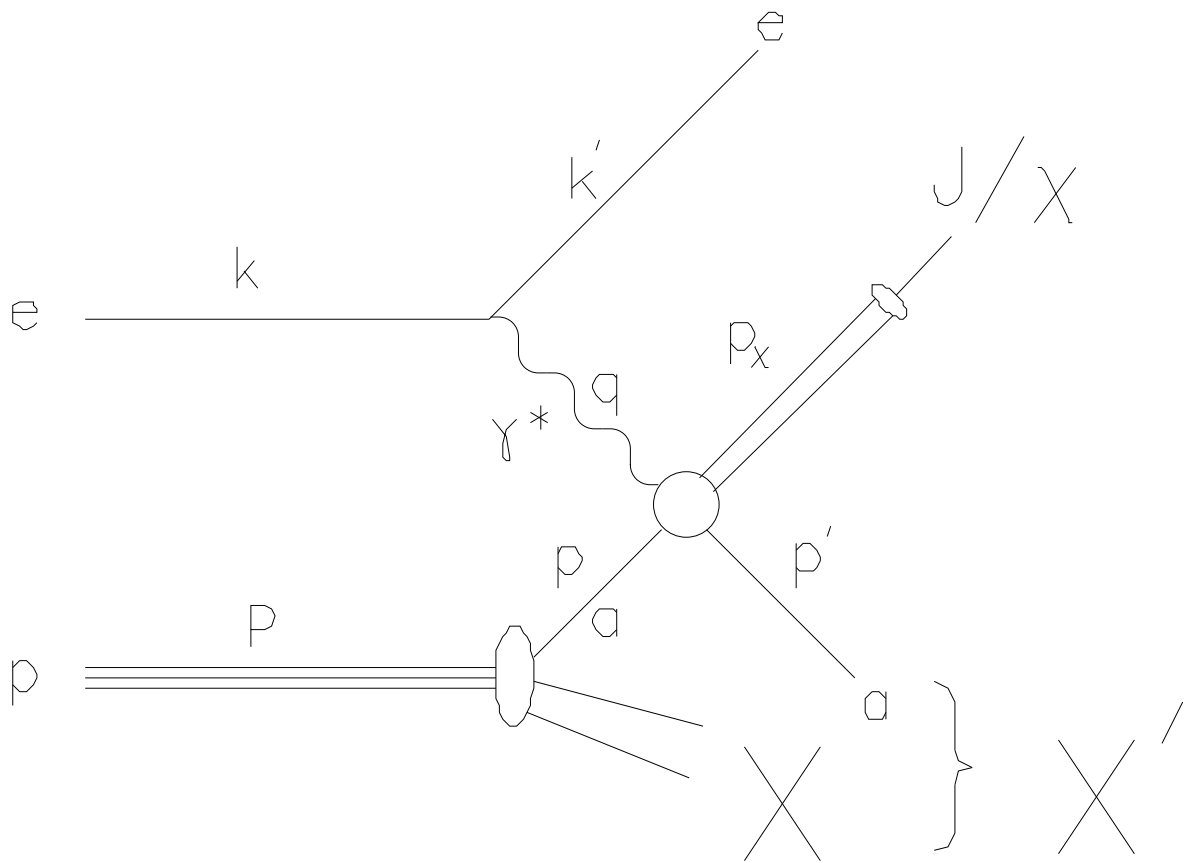,width=16cm,bbllx=222pt,bblly=194pt,bburx=306pt,%
bbury=257pt}
}
\caption{Schematic representation of $e+p\to e+J/\psi+j+X$ explaining the
four-momentum assignments.}
\label{fig:kin}
\end{center}
\end{figure}

In this section, we present our analytic results for the cross section of
$e+p\to e+J/\psi+j+X$ in DIS.
We work at LO in the parton model of QCD with $n_f=3$ active quark flavours
and employ the NRQCD factorization formalism \cite{bbl} to describe the
formation of the $J/\psi$ meson.
We start by defining the kinematics.
As indicated in Fig.~\ref{fig:kin}, we denote the four-momenta of the incoming
lepton and proton and the outgoing lepton, $J/\psi$ meson, and jet by $k$,
$P$, $k^\prime$, $p_\psi$, and $p^\prime$, respectively.
The parton struck by the virtual photon carries four-momentum $p=xP$.
We neglect the masses of the proton, lepton, and light quarks, call the one
of the $J/\psi$ meson $M_\psi$, and take the charm-quark mass to be
$m_c=M_\psi/2$.
In our approximation, the proton remnant $X$ has zero invariant mass,
$M_X^2=(P-p)^2=0$.
The CM energy square of the $ep$ collision is $S=(k+P)^2$.
The virtual photon has four-momentum $q=k-k^\prime$, and it is customary
\cite{pdg} to define $Q^2=-q^2>0$ and $y=q\cdot P/k\cdot P$, which measures
the relative lepton energy loss in the proton rest frame.
The inelasticity variable, which was already mentioned in the Introduction, is
defined as $z=p_\psi\cdot P/q\cdot P$.
The system $X^\prime$ consisting of the jet $j$ and the proton remnant $X$ has
invariant mass square $M_{X^\prime}^2=(q+P-p_\psi)^2=(1-x)y(1-z)S$.
Other frequently employed variables \cite{pdg} are Bjorken's variable
$x_B=Q^2/(2q\cdot P)=Q^2/(yS)$ and the $\gamma^\star p$ CM energy square
$W^2=(q+P)^2=yS-Q^2$.
As usual, we define the partonic Mandelstam variables as
$\hat s=(q+p)^2=xyS-Q^2$, $\hat t=(q-p_\psi)^2=-xy(1-z)S$, and
$\hat u=(p-p_\psi)^2=M_\psi^2-xyzS$.
By four-momentum conservation, we have $\hat s+\hat t+\hat u=M_\psi^2-Q^2$.
In the $\gamma^\star p$ CM frame, the $J/\psi$ meson has transverse momentum
and rapidity
\begin{eqnarray}
p_T^\star&=&\frac{\sqrt{\hat t\left(\hat s\hat u+Q^2M_\psi^2\right)}}
{\hat s+Q^2},
\\
y_\psi^\star&=&\frac{1}{2}\ln\frac{\hat s\left(M_\psi^2-\hat u\right)}
{\hat s\left(M_\psi^2-\hat t\right)+Q^2M_\psi^2}
+\frac{1}{2}\ln\frac{W^2}{\hat s},
\label{eq:ycms}
\end{eqnarray}
respectively.
Here and in the following, we denote the quantities referring to the
$\gamma^\star p$ CM frame by an asterisk.
The second term on the right-hand side of Eq.~(\ref{eq:ycms}) originates from 
the Lorentz boost from the $\gamma^\star a$ CM frame to the $\gamma^\star p$
one.
Here, $y_\psi^\star$ is taken to be positive in the direction of the
three-momentum of the virtual photon, in accordance with the convention of
Refs.~\cite{mey,h1,mor}.

In the parton model, the proton is characterized by its PDFs $f_{a/p}(x,M)$,
and, at LO, an outgoing parton may be identified with a jet.
Thus, we have
\begin{equation}
d\sigma(e+p\to e+J/\psi+j+X)
=\int_0^1dx\sum_af_{a/p}(x,M)d\sigma(e+a\to e+J/\psi+a),
\label{eq:par}
\end{equation}
where $a=g,u,\overline{u},d,\overline{d},s,\overline{s}$.
Furthermore, according to the NRQCD factorization formalism, we have
\begin{equation}
d\sigma(e+a\to e+J/\psi+a)
=\sum_n\left\langle{\cal O}^\psi[n]\right\rangle
d\sigma(e+a\to e+c\overline{c}[n]+a),
\label{eq:fac}
\end{equation}
where, to LO in $v$,
$n={}^3\!S_1^{(1)},{}^1\!S_0^{(8)},{}^3\!S_1^{(8)},{}^3\!P_J^{(8)}$.

Decomposing the transition-matrix element of the partonic subprocess
$e+a\to e+c\overline{c}[n]+a$ into a leptonic part,
${\cal T}^\mu(e\to e+\gamma^\star)
=-(e/q^2)\overline{u}(k^\prime)\gamma^\mu u(k)$,
and a hadronic one, ${\cal T}^\mu(\gamma^\star+a\to c\overline{c}[n]+a)$, from
which the virtual-photon leg is amputated, we can write its cross section as
\begin{equation}
d\sigma(e+a\to e+c\overline{c}[n]+a)
=\frac{1}{2xS}\,\frac{1}{4N_a}\,\frac{e^2}{(q^2)^2}
\tr(\not k\gamma^\nu\not k^\prime\gamma^\mu)H_{\mu\nu}
d{\mathrm PS}_3(k+p;k^\prime,p_\psi,p^\prime),
\label{eq:dec}
\end{equation}
where $N_g=(N_c^2-1)$ and $N_q=N_{\overline{q}}=N_c=3$ are the colour
multiplicities of the partons $a$, $e$ is the electron charge magnitude, and
the hadronic tensor $H^{\mu\nu}$ is obtained by summing the absolute square of
${\cal T}^\mu(\gamma^\star+a\to c\overline{c}[n]+a)$ over the spin and colour
states of the incoming and outgoing partons $a$.
Here and in the following, we employ the Lorentz-invariant phase-space measure
\begin{equation}
d{\mathrm PS}_n(p;p_1,\ldots,p_n)
=(2\pi)^4\delta^{(4)}\left(p-\sum_{i=1}^np_i\right)\prod_{i=1}^n
\frac{d^3p_i}{(2\pi)^32p_i^0}.
\end{equation}
The first factor in Eq.~(\ref{eq:dec}) stems from the flux and the second one
from the average over the spin and colour states of the incoming particles.
Integrating over the azimuthal angle of the outgoing lepton, we may simplify
Eq.~(\ref{eq:dec}) to become
\begin{equation}
d\sigma(e+a\to e+c\overline{c}[n]+a)=\frac{1}{2xS}\,\frac{1}{4N_a}\,
\frac{\alpha}{2\pi}L^{\mu\nu}H_{\mu\nu}\frac{dy}{y}\,\frac{dQ^2}{Q^2}
d{\mathrm PS}_2(q+p;p_\psi,p^\prime),
\label{eq:red}
\end{equation}
where $\alpha=e^2/(4\pi)$ is Sommerfeld's fine-structure constant and
\cite{gra}
\begin{equation}
L^{\mu\nu}=\frac{1+(1-y)^2}{y}\epsilon_T^{\mu\nu}
-\frac{4(1-y)}{y}\epsilon_L^{\mu\nu},
\label{eq:lep}
\end{equation}
with
\begin{eqnarray}
\epsilon_T^{\mu\nu}&=&-g^{\mu\nu}+\frac{1}{q\cdot p}(q^\mu p^\nu+p^\mu q^\nu)
-\frac{q^2}{(q\cdot p)^2}p^\mu p^\nu,
\nonumber\\
\epsilon_L^{\mu\nu}&=&\frac{1}{q^2}\left(q-\frac{q^2}{q\cdot p}p\right)^\mu
\left(q-\frac{q^2}{q\cdot p}p\right)^\nu,
\end{eqnarray}
is the leptonic tensor.
We have $q_\mu\epsilon_T^{\mu\nu}=q_\mu\epsilon_L^{\mu\nu}=0$,
$\epsilon_{T\mu}^\mu=-2$, and $\epsilon_{L\mu}^\mu=-1$.
Furthermore,
\begin{equation}
\epsilon^{\mu\nu}=\epsilon_T^{\mu\nu}+\epsilon_L^{\mu\nu}
=-g^{\mu\nu}+\frac{q^\mu q^\nu}{q^2}
\end{equation}
is the polarization tensor of an unpolarized spin-one boson with mass $q^2$.
In the $\gamma^\star a$ CM frame, where $q^\mu=(q^0,0,0,q^3)$ and
$p^\mu=(q^3,0,0,-q^3)$, we have $\epsilon_T^{\mu\nu}=\diag(0,1,1,0)$ and
$\epsilon_L^{\mu\nu}=(1/q^2)(q^3,0,0,q^0)^\mu(q^3,0,0,q^0)^\nu$, so that
$\epsilon_T^{\mu\nu}$ and $\epsilon_L^{\mu\nu}$ refer to transverse and
longitudinal polarization, as indicated by their subscripts.
Since the hadronic current is conserved in QED, we have $q_\mu H^{\mu\nu}=0$,
which leads to a further simplification as Eq.~(\ref{eq:lep}) is inserted
in Eq.~(\ref{eq:red}).

It is interesting to study the photoproduction limit, by taking $Q^2\to0$ in
Eq.~(\ref{eq:red}).
This provides us with a powerful check for our results by relating them to
well-known results in the literature \cite{ber,bkv,ko}.
The differential cross section of the partonic process
$\gamma+a\to c\overline{c}[n]+a$ reads
\begin{equation}
d\sigma(\gamma+a\to c\overline{c}[n]+a)=\frac{1}{2\hat s}\,\frac{1}{4N_a}\,
(-g^{\mu\nu})H_{\mu\nu}|_{Q^2=0}d{\mathrm PS}_2(q+p;p_\psi,p^\prime).
\label{eq:pho}
\end{equation}
Comparing Eqs.~(\ref{eq:red}) and (\ref{eq:pho}), we thus obtain the master 
formula
\begin{equation}
\lim_{Q^2\to0}\frac{Q^2d^2\sigma}{dy\,dQ^2}(e+a\to e+c\overline{c}[n]+a)
=\frac{\alpha}{2\pi}\,\frac{1+(1-y)^2}{y}
\sigma(\gamma+a\to c\overline{c}[n]+a).
\label{eq:mas}
\end{equation}
A similar relationship between the DIS process $e+p\to e+J/\psi+X$ and the
photoproduction one $\gamma+p\to J/\psi+X$ may be found in Eq.~(4) of
Ref.~\cite{fle}.

We evaluate the cross sections of the relevant partonic subprocesses
$e+a\to e+c\overline{c}[n]+a$ from Eq.~(\ref{eq:red}) applying the
covariant-projector method of Ref.~\cite{pet}.
Our results can be written in the form
\begin{equation}
\frac{d^3\sigma}{dy\,dQ^2\,d\hat t}(e+a\to e+c\overline{c}[n]+a)
=\frac{\alpha}{2\pi}F_a[n]\left[\frac{1+(1-y)^2}{yQ^2}T_a[n]
-\frac{4(1-y)}{y}L_a[n]\right],
\label{eq:res}
\end{equation}
where $F_a[n]$, $T_a[n]$, and $L_a[n]$ are functions of $\hat s$, $\hat t$,
$\hat u$, and $Q^2$, which are listed in the Appendix.
They are finite for $Q^2=0$.
We combined the results for $n={}^3\!P_J^{(8)}$, with $J=0,1,2$, exploiting
the multiplicity relation
\begin{equation}
\left\langle{\cal O}^\psi\left[{}^3\!P_J^{(8)}\right]\right\rangle
=(2J+1)\left\langle{\cal O}^\psi\left[{}^3\!P_0^{(8)}\right]\right\rangle,
\end{equation}
which follows to LO in $v$ from heavy-quark spin symmetry.
We recover the well-known cross sections of the corresponding CS \cite{ber} 
and CO \cite{bkv,ko} processes of photoproduction by inserting
Eq.~(\ref{eq:res}) in Eq.~(\ref{eq:mas}), as
\begin{equation}
\frac{d\sigma}{d\hat t}(\gamma+a\to c\overline{c}[n]+a)=F_a[n]T_a[n]|_{Q^2=0}.
\end{equation}

At this point, we should compare our analytic results with the literature.
Formulas for the cross section of the CS partonic subprocess
$e+g\to e+c\overline{c}\left[{}^3\!S_1^{(1)}\right]+g$ may be found in
Refs.~\cite{koe,gui,mer,kru,yua}.
In Ref.~\cite{yua}, also cross section formulas for the CO partonic 
subprocesses $e+a\to e+c\overline{c}[n]+a$ with $a=g,q,\overline{q}$ and
$n={}^1\!S_0^{(8)},{}^3\!S_1^{(8)},{}^3\!P_J^{(8)}$ are listed.
We agree with the result for
$e+g\to e+c\overline{c}\left[{}^1\!S_0^{(8)}\right]+g$ \cite{yua}, but we
disagree with all the other results.
In particular, the results of Refs.~\cite{koe,mer} and the residual results
of Ref.~\cite{yua} fail to reproduce the well-established formulas of
Refs.~\cite{ber,bkv,ko} in the photoproduction limit.
We also remark that Eqs.~(A21) and (A35)--(A37) of Ref.~\cite{yua} suffer from
mass-dimensional inconsistencies.
Furthermore, we only find agreement with the result in Eq.~(4) of
Ref.~\cite{gui} if we flip the overall sign of $L_a[n]$ in Eq.~(\ref{eq:res}).
However, this causes the cross section in Eq.~(\ref{eq:res}) to turn negative
in certain regions of phase space, e.g.\ at $y=0.5$, $Q^2=25$~GeV${}^2$,
$\hat s=100$~GeV${}^2$, and $\hat t=-10$~GeV${}^2$.
Similarly, we only agree with the result in Eq.~(2.49) of Ref.~\cite{kru} if
we include an overall factor of $4\pi\alpha$ on the right-hand side of
Eq.~(\ref{eq:res}) and halve $L_a[n]$.
As for the cross sections of the CO partonic subprocesses
$e+g\to e+c\overline{c}[n]$ with $n={}^1\!S_0^{(8)},{}^3\!P_J^{(8)}$, which do
not enter our analysis, we agree with Eq.~(3) of Ref.~\cite{fle}.

The cross sections of $e+g\to e+c\overline{c}[n]+g$ with
$n={}^1\!S_0^{(8)},{}^3\!P_J^{(8)}$ exhibit collinear singularities in the
limit $\hat t\to0$.
According to the factorization theorem of the QCD-improved parton model, the
limiting expressions must coincide with the $e+g\to e+c\overline{c}[n]$ cross
sections \cite{fle} multiplied by the spacelike $g\to g$ splitting functions.
This provides another nontrivial check for our results, and, among other
things, this fixes the overall factor of $L_a[n]$.

Inserting Eq.~(\ref{eq:fac}) in Eq.~(\ref{eq:par}) and including the maximum
boundaries of the integrations over $x$ and $\hat t$, we obtain
\begin{eqnarray}
\lefteqn{\frac{d^2\sigma}{dy\,dQ^2}(e+p\to e+J/\psi+j+X)
=\int_{(Q^2+M_\psi^2)/(yS)}^1dx
\int_{-(\hat s+Q^2)(\hat s-M_\psi^2)/\hat s}^0d\hat t}
\nonumber\\
&&{}\times
\sum_af_{a/p}(x,M)\sum_n\left\langle{\cal O}^\psi[n]\right\rangle
\frac{d^3\sigma}{dy\,dQ^2\,d\hat t}(e+a\to e+c\overline{c}[n]+a),
\label{eq:dif}
\end{eqnarray}
where $\left(d^3\sigma/dy\,dQ^2\,d\hat t\right)(e+a\to e+c\overline{c}[n]+a)$
is given by Eq.~(\ref{eq:res}).
The kinematically allowed ranges of $y$ and $Q^2$ are $M_\psi^2/S<y<1$ and
$0<Q^2<yS-M_\psi^2$, respectively.
In order to avoid the collinear singularities mentioned above, we need to 
reduce the upper boundary of the $\hat t$ integration in Eq.~(\ref{eq:dif}).
This cut-off should, of course, depend on variables that can be controlled
experimentally.
It is convenient to introduce an upper cut-off, below unity, on the
inelasticity variable $z$.
As explained in the Introduction, such a cut-off also suppresses the
diffractive background.
The distributions in $y$ and $Q^2$ can be evaluated from Eq.~(\ref{eq:dif}) as
it stands.
It is also straightforward to obtain the distributions in $z$, $x_B$, $W$,
$p_T^\star$, and $y_\psi^\star$, by accordingly redefining and reordering the
integration variables in Eq.~(\ref{eq:dif}).
The distribution in the $J/\psi$ azimuthal angle $\phi^\star$ in the
$\gamma^\star p$ CM frame is constant.

The evaluation of the distributions in the $J/\psi$ transverse momentum $p_T$,
rapidity $y_\psi$, and azimuthal angle $\phi$ in the HERA laboratory frame is
somewhat more involved.
Choosing a suitable coordinate system in the $\gamma^\star p$ CM frame, we
have
\begin{eqnarray}
(k^\star)^\mu&=&\frac{S-Q^2}{2W}\left(
\begin{array}{c}
1\\
\sin\psi^\star\\
0\\
\cos\psi^\star
\end{array}
\right),
\qquad
(k^{\prime\star})^\mu=\frac{S-W^2}{2W}\left(
\begin{array}{c}
1\\
\sin\theta^\star\\
0\\
\cos\theta^\star
\end{array}
\right),
\nonumber\\
(q^\star)^\mu&=&\frac{1}{2W}\left(
\begin{array}{c}
W^2-Q^2\\
0\\
0\\
W^2+Q^2
\end{array}
\right),
\qquad
(P^\star)^\mu=\frac{W^2+Q^2}{2W}\left(
\begin{array}{c}
1\\
0\\
0\\
-1
\end{array}
\right),
\nonumber\\
(p_\psi^\star)^\mu&=&\left(
\begin{array}{c}
m_T^\star\cosh y_\psi^\star\\
p_T^\star\cos\phi^\star\\
p_T^\star\sin\phi^\star\\
m_T^\star\sinh y_\psi^\star
\end{array}
\right),
\end{eqnarray}
where $\cos\psi^\star=2SW^2/[(S-Q^2)(W^2+Q^2)]-1$,
$\cos\theta^\star=1-2SQ^2/[(S-W^2)(W^2+Q^2)]$, and
$m_T^\star=\sqrt{M_\psi^2+\left(p_T^\star\right)^2}$ is the $J/\psi$
transverse mass.
On the other hand, in the laboratory frame, we have
\begin{eqnarray}
k^\mu&=&E_e\left(
\begin{array}{c}
1\\
0\\
0\\
1
\end{array}
\right),
\qquad
(k^\prime)^\mu=E_e^\prime\left(
\begin{array}{c}
1\\
\sin\theta\\
0\\
\cos\theta
\end{array}
\right),
\nonumber\\
q^\mu&=&\left(
\begin{array}{c}
q^0\\
-q\sin\psi\\
0\\
q\cos\psi
\end{array}
\right),
\qquad
P^\mu=E_p\left(
\begin{array}{c}
1\\
0\\
0\\
-1
\end{array}
\right),
\nonumber\\
p_\psi^\mu&=&\left(
\begin{array}{c}
m_T\cosh y_\psi\\
p_T\cos\phi\\
p_T\sin\phi\\
m_T\sinh y_\psi
\end{array}
\right),
\end{eqnarray}
where $E_e$ and $E_p$ are the lepton and proton energies, respectively,
$E_e^\prime=[(S-W^2-Q^2)/E_p+Q^2/E_e]/4$,
$\cos\theta=1-Q^2/\left(2E_eE_e^\prime\right)$,
$q^0=[(W^2+Q^2)/E_p-Q^2/E_e]/4$, $q=\sqrt{Q^2+(q^0)^2}$,
$\cos\psi=[(W^2+Q^2)/E_p+Q^2/E_e]/(4q)$, and $m_T=\sqrt{M_\psi^2+p_T^2}$ is
the $J/\psi$ transverse mass.
Notice that $y_\psi$ is taken to be positive in the direction of the
three-momentum of the incoming lepton.
Without loss of generality, we may require that $0\le\psi^\star,\psi\le\pi$,
for, otherwise, we can achieve this by rotating the respective coordinate
systems by $180^\circ$ around the $z$ axis.
We can then evaluate $p_T$, $y_\psi$, and $\phi$ from $p_T^\star$,
$y_\psi^\star$, and $\phi^\star$ as
\begin{eqnarray}
p_T&=&\sqrt{\left(p_T^\star\right)^2+A\left(A-2p_T^\star\cos\phi^\star\right)},
\label{eq:ptlab}
\\
y_\psi&=&y_\psi^\star+\ln\frac{(W^2+Q^2)m_T^\star}{\sqrt SWm_T}
+\frac{1}{2}\ln\frac{E_e}{E_p},
\label{eq:ylab}
\\
\cos\phi&=&\frac{p_T^\star\cos\phi^\star-A}{p_T},
\label{eq:philab}
\end{eqnarray}
where 
\begin{equation}
A=\frac{m_T^\star\exp(y_\psi^\star)\sin\psi^\star}{1+\cos\psi^\star}
=\sqrt{\frac{Q^2(S-W^2-Q^2)}{SW^2}}m_T^\star\exp(y_\psi^\star).
\label{eq:a}
\end{equation}
The third term on the right-hand side of Eq.~(\ref{eq:ylab}) stems from the
Lorentz boost from the $ep$ CM frame to the laboratory one.
Since $p_T$, $y_\psi$, and $\phi$ depend on $\phi^\star$, the integration over
$\phi^\star$ in Eq.~(\ref{eq:dif}) is no longer trivial, and we need to insert
the symbolic factor $(1/2\pi)\int_0^{2\pi}d\phi^\star$ on the right-hand side
of that equation.

For future applications, we also present compact formulas that allow us to
determine $p_T^\star$, $y_\psi^\star$, and $\phi^\star$, once $p_T$, $y_\psi$,
and $\phi$ are given.
In fact, Eqs.~(\ref{eq:ptlab}) and (\ref{eq:philab}) can be straightforwardly
inverted by observing that the quantity $A$ defined in Eq.~(\ref{eq:a}) can be
expressed in terms of $m_T$ and $y_\psi$ by substituting Eq.~(\ref{eq:ylab}),
the result being
\begin{equation}
A=\frac{m_T\exp(y_\psi)}{W^2+Q^2}\sqrt{\frac{E_p}{E_e}Q^2(S-W^2-Q^2)}.
\end{equation}
Having obtained $p_T^\star$, we can then evaluate $y_\psi^\star$ from 
Eq.~(\ref{eq:ylab}).
For the reader's convenience, we collect the relevant formulas here:
\begin{eqnarray}
p_T^\star&=&\sqrt{p_T^2+A(A+2p_T\cos\phi)},
\nonumber\\
y_\psi^\star&=&y_\psi+\ln\frac{\sqrt SWm_T}{(W^2+Q^2)m_T^\star}
+\frac{1}{2}\ln\frac{E_p}{E_e},
\nonumber\\
\cos\phi^\star&=&\frac{p_T\cos\phi+A}{p_T^\star}.
\end{eqnarray}

\section{Numerical results}
\label{sec:three}

We are now in a position to present our numerical results.
We first describe our theoretical input and the kinematic conditions.
We use $m_c=(1.5\pm0.1)$~GeV, $\alpha=1/137.036$, and the LO formula for 
$\alpha_s^{(n_f)}(\mu)$ with $n_f=3$ active quark flavours \cite{pdg}.
As for the proton PDFs, we employ the LO set by Martin, Roberts, Stirling, and 
Thorne (MRST98LO) \cite{mrst}, with asymptotic scale parameter
$\Lambda^{(4)}=174$~MeV, as our default and the LO set by the CTEQ
Collaboration (CTEQ5L) \cite{cteq}, with $\Lambda^{(4)}=192$~MeV, for
comparison.
The corresponding values of $\Lambda^{(3)}$ are 204~MeV and 224~MeV, 
respectively.
We choose the renormalization and factorization scales to be
$\mu=M=\xi\sqrt{Q^2+M_\psi^2}$ and vary the scale parameter $\xi$ between
1/2 and 2 abound the default value 1.
We adopt the NRQCD MEs from Table~I of Ref.~\cite{bkl}.
Specifically, they read
$\left\langle{\cal O}^\psi\left[{}^3\!S_1^{(1)}\right]\right\rangle
=(1.3\pm0.1)$~GeV${}^3$,
$\left\langle{\cal O}^\psi\left[{}^3\!S_1^{(8)}\right]\right\rangle
=(4.4\pm0.7)\times10^{-3}$~GeV${}^3$, and
$M_{3.4}^\psi=(8.7\pm0.9)\times10^{-2}$~GeV${}^3$ for set MRST98LO, where
\begin{equation}
M_r^\psi=\left\langle{\cal O}^\psi\left({}^1\!S_0^{(8)}\right)\right\rangle
+\frac{r}{m_c^2}
\left\langle{\cal O}^\psi\left({}^3\!P_0^{(8)}\right)\right\rangle.
\label{eq:mr}
\end{equation}
The corresponding values for set CTEQ5L are
$(1.4\pm0.1)$~GeV${}^3$, $(3.9\pm0.7)\times10^{-3}$~GeV${}^3$, and
$(6.6\pm0.7)\times10^{-2}$~GeV${}^3$, respectively.
Since Eq.~(\ref{eq:dif}) is sensitive to a different linear combination of
$\left\langle{\cal O}^\psi\left({}^1\!S_0^{(8)}\right)\right\rangle$ and
$\left\langle{\cal O}^\psi\left({}^3\!P_0^{(8)}\right)\right\rangle$ than 
appears in Eq.~(\ref{eq:mr}), we write
$\left\langle{\cal O}^\psi\left({}^1\!S_0^{(8)}\right)\right\rangle=\kappa
M_r^\psi$
and
$\left\langle{\cal O}^\psi\left({}^3\!P_0^{(8)}\right)\right\rangle=(1-\kappa)
\left(m_c^2/r\right)M_r^\psi$ and vary $\kappa$ between 0 and 1 around the
default value 1/2.
In order to estimate the theoretical uncertainties in our predictions, we
vary the unphysical parameters $\xi$ and $\kappa$ as indicated above, take
into account the experimental errors on $m_c$ and the default MEs, and switch
from our default PDF set to the CTEQ5L one, properly adjusting $\Lambda^{(3)}$
and the MEs.
We then combine the individual shifts in quadrature.

The H1 data on $J/\psi$ inclusive and inelastic production in DIS
\cite{mey,h1} were taken in collisions of positrons with $E_e=27.5$~GeV and
protons with $E_p=820$~GeV in the HERA laboratory frame, so that
$\sqrt S=2\sqrt{E_eE_p}=300$~GeV, and they refer to the kinematic region
defined by $2<Q^2<80$~GeV${}^2$, $40<W<180$~GeV, and $z>0.2$.
In Ref.~\cite{h1}, the acceptance cut $M_{X^\prime}>10$~GeV on the invariant
mass of the hadronic system $X^\prime$ produced in association with the
$J/\psi$ meson was imposed in order to exclude the contribution of $J/\psi$
elastic production, which, to a large extent, is due to diffractive processes.
Notice that this cut allows for $z$ to be as large as
$z_{\mathrm max}=1-M_{X^\prime,{\mathrm min}}^2/\left(W_{\mathrm max}^2
-M_\psi^2\right)\approx0.997$ and for $\left|\hat t\right|$ to be as small as
$\left|\hat t\right|_{\mathrm min}=(1-z_{\mathrm max})\left(Q_{\mathrm min}
+M_\psi^2\right)\approx0.040$~GeV${}^2$.
A more conservative way to eliminate the domain of $J/\psi$ elastic production
is to directly impose the cut $z<0.9$ \cite{mey}, which ensures that
$\left|\hat t\right|$ is in excess of
$\left|\hat t\right|_{\mathrm min}\approx1.3$~GeV${}^2$.
At the same time, a sufficiently large lower bound on $\left|\hat t\right|$ is
requisite in order to screen Eq.~(\ref{eq:dif}) from the collinear
singularities in the ${}^1\!S_0^{(8)}$ and ${}^3\!P_J^{(8)}$ channels,
mentioned in Section~\ref{sec:two}, and, thus, to keep our theoretical
predictions perturbatively stable.
In the following, we, therefore, use the H1 data with the cut $z<0.9$, as
presented in Ref.~\cite{mey}, for comparisons.

\begin{figure}[ht]
  \begin{center}
    \setlength{\unitlength}{1cm}
    \begin{picture}(15,10)
      \setlength{\unitlength}{1cm}
      \put(0.5,0){\includegraphics[width=15cm,height=10cm]{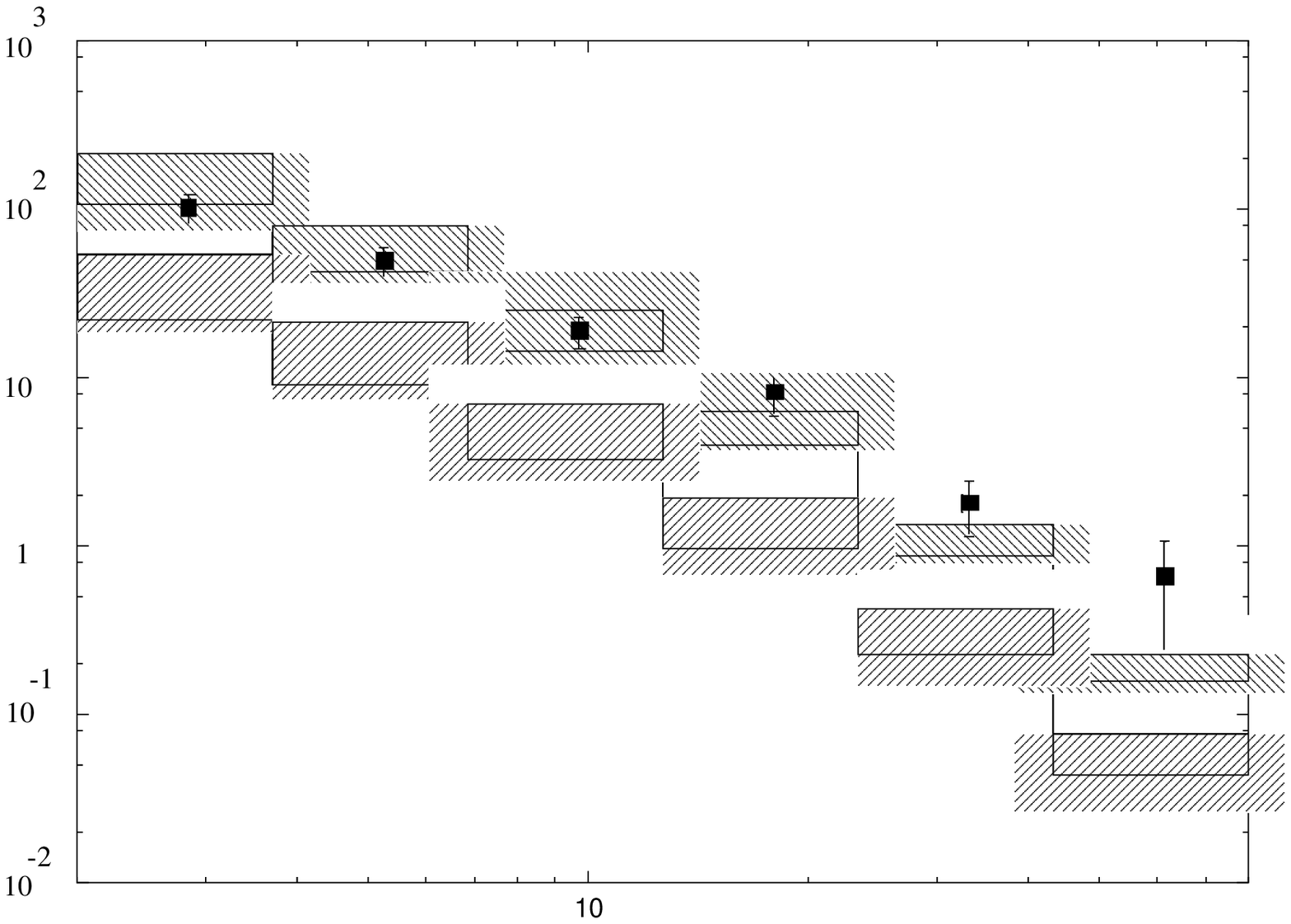}}
      \put(-1.5,5){$\displaystyle\frac{d\sigma}{dQ^2}$}
      \put(-1.5,4){[pb/GeV${}^2$]}
      \put(7.5,-0.5){$Q^2$ [GeV${}^2$]}
    \end{picture}
  \end{center}
\caption{The $Q^2$ distribution of $J/\psi$ inclusive production in $ep$ DIS
measured with the H1 detector \protect\cite{mey} is compared with the LO NRQCD
(upper histogram) and CSM (lower histogram) predictions.}
\label{fig:Q2}
\end{figure}

\begin{figure}[ht]
  \begin{center}
    \setlength{\unitlength}{1cm}
    \begin{picture}(15,10)
      \setlength{\unitlength}{1cm}
      \put(0.5,0){\includegraphics[width=15cm,height=10cm]{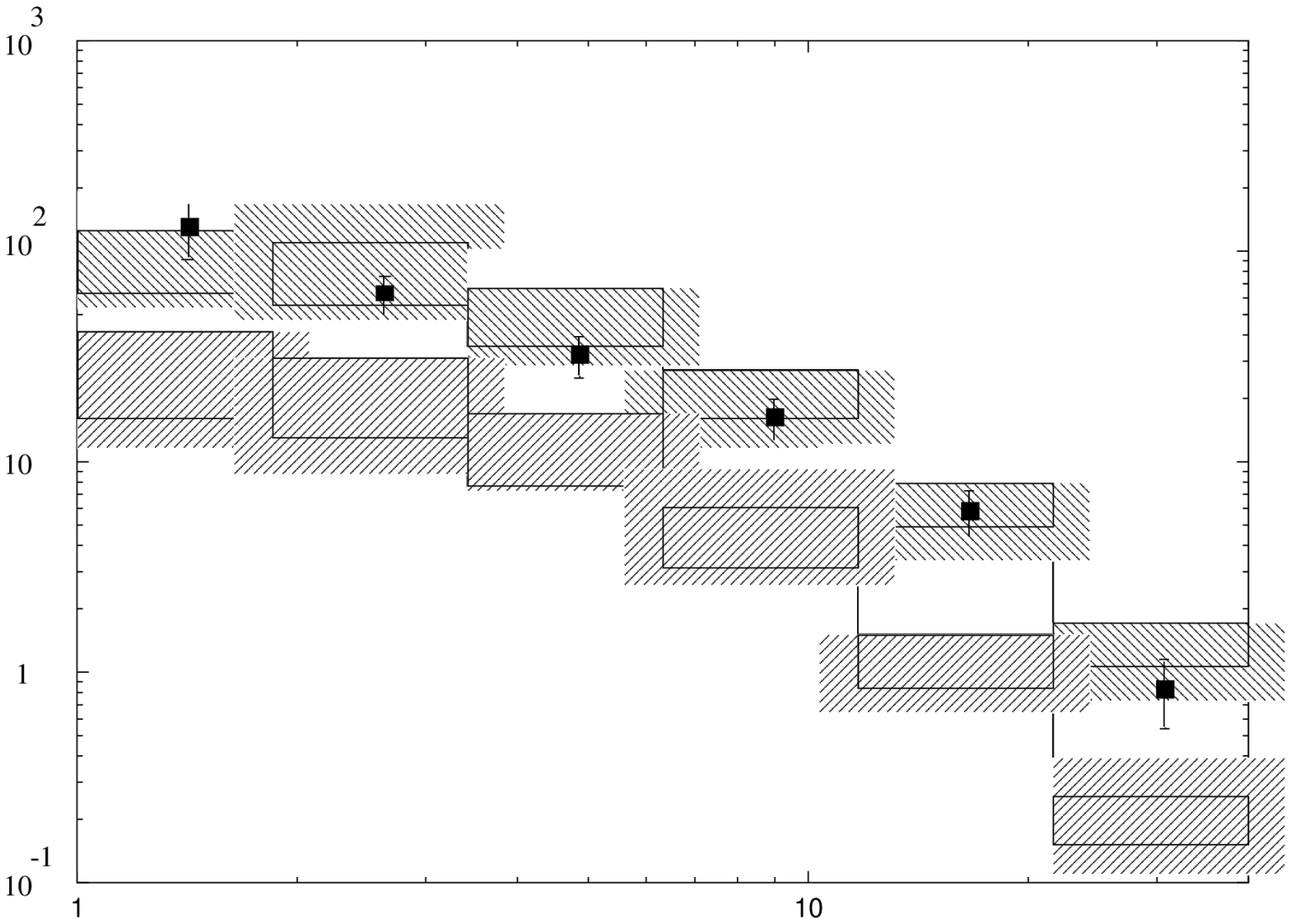}}
      \put(-1.5,5){$\displaystyle\frac{d\sigma}{dp_T^2}$}
      \put(-1.5,4){[pb/GeV${}^2$]}
      \put(7.5,-0.5){$p_T^2$ [GeV${}^2$]}
    \end{picture}
  \end{center}
\caption{The $p_T^2$ distribution of $J/\psi$ inclusive production in $ep$ DIS
measured with the H1 detector \protect\cite{mey} is compared with the LO NRQCD
(upper histogram) and CSM (lower histogram) predictions.}
\label{fig:pT2Lab}
\end{figure}

\begin{figure}[ht]
  \begin{center}
    \setlength{\unitlength}{1cm}
    \begin{picture}(15,10)
      \setlength{\unitlength}{1cm}
      \put(0.5,0){\includegraphics[width=15cm,height=10cm]{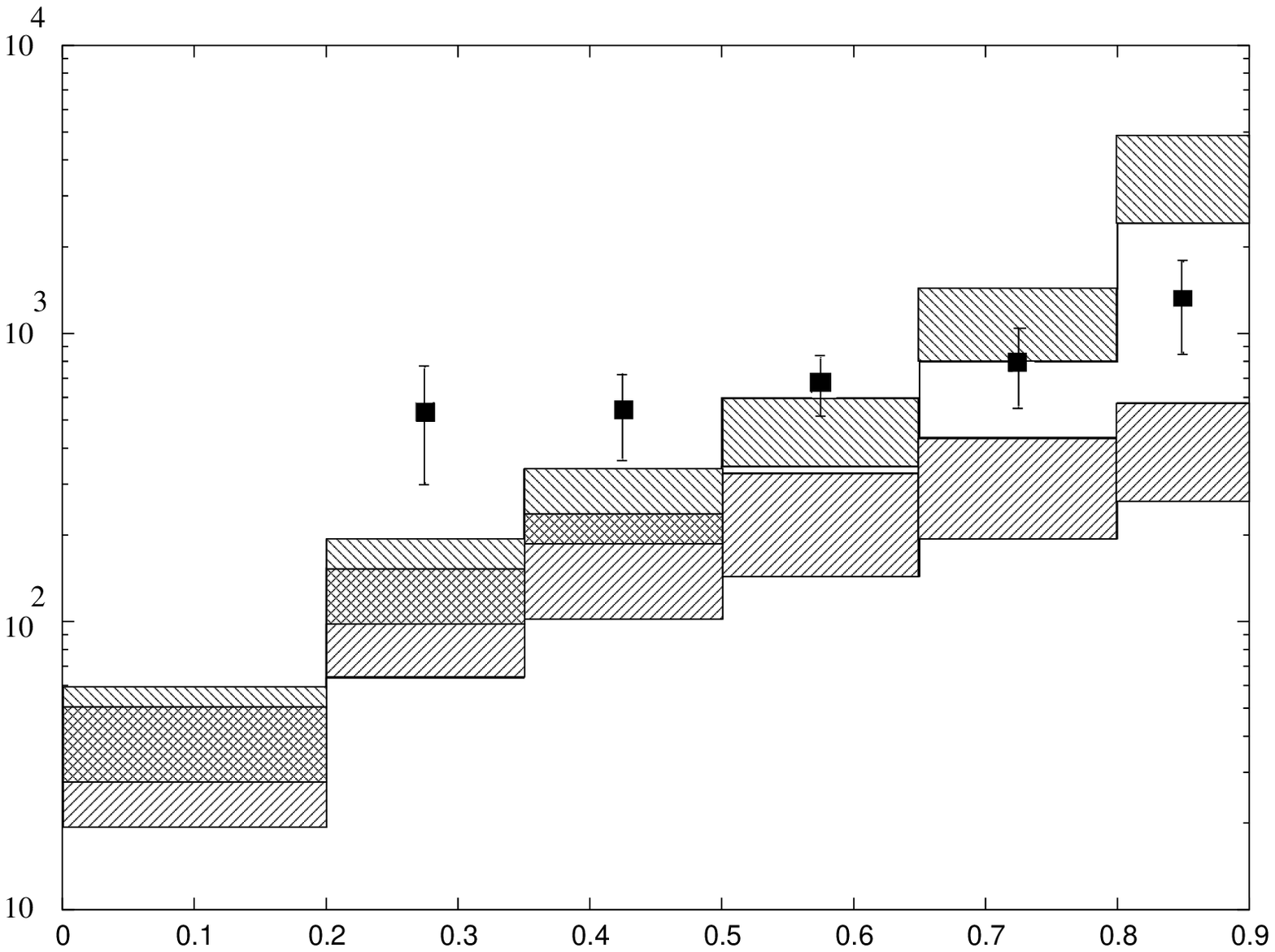}}
      \put(0,5){$\displaystyle\frac{d\sigma}{dz}$}
      \put(0,4){[pb]}
      \put(8.5,-0.5){$z$}
    \end{picture}
  \end{center}
\caption{The $z$ distribution of $J/\psi$ inclusive production in $ep$ DIS
measured with the H1 detector \protect\cite{mey} is compared with the LO NRQCD
(upper histogram) and CSM (lower histogram) predictions.}
\label{fig:z}
\end{figure}

\begin{figure}[ht]
  \begin{center}
    \setlength{\unitlength}{1cm}
    \begin{picture}(15,10)
      \setlength{\unitlength}{1cm}
      \put(0.5,0){\includegraphics[width=15cm,height=10cm]{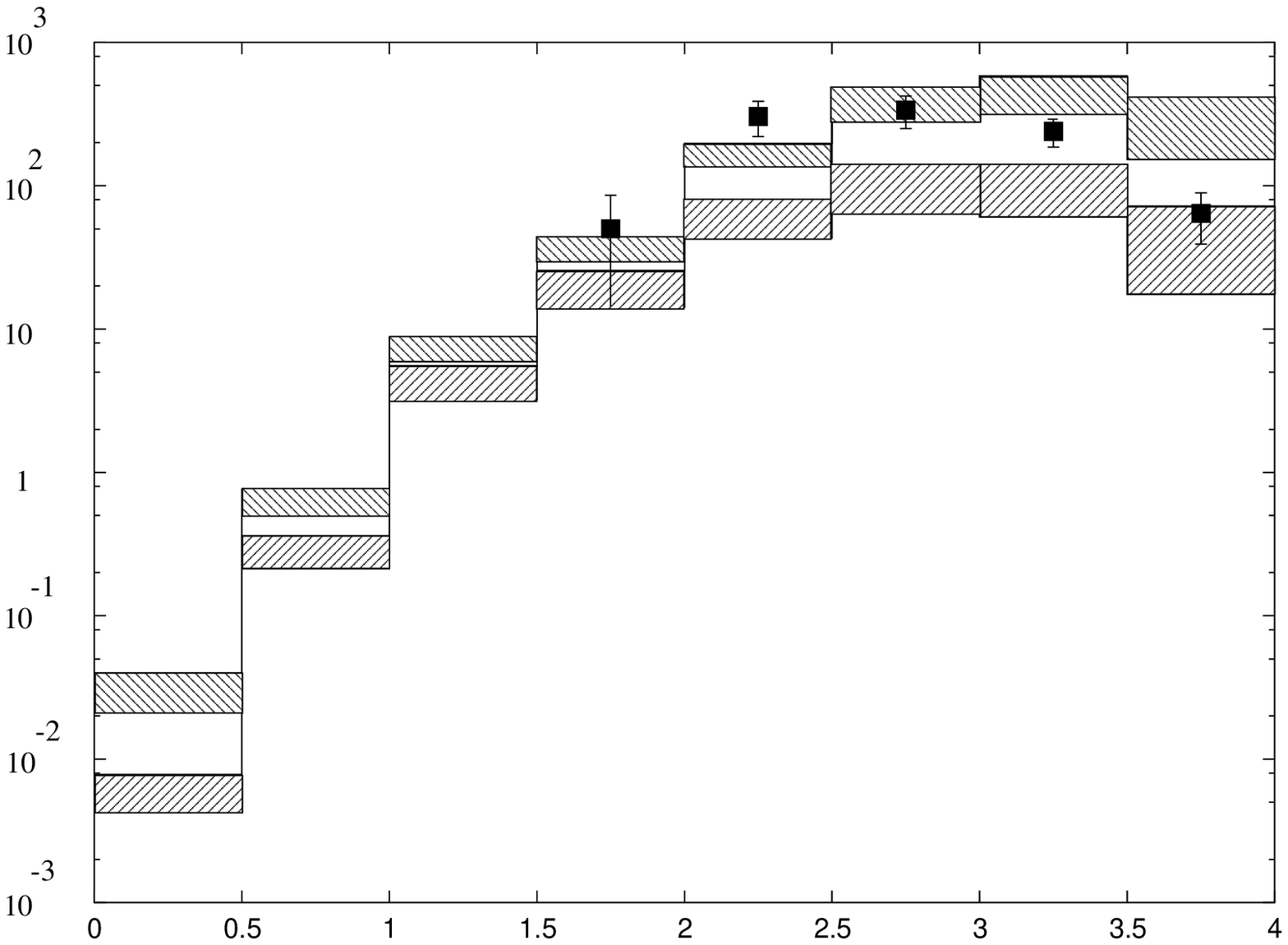}}
      \put(0,5){$\displaystyle\frac{d\sigma}{dy^\star}$}
      \put(0,4){[pb]}
      \put(8.5,-0.5){$y^\star$}
    \end{picture}
  \end{center}
\caption{The $y^\star$ distribution of $J/\psi$ inclusive production in $ep$
DIS measured with the H1 detector \protect\cite{mey} is compared with the LO
NRQCD (upper histogram) and CSM (lower histogram) predictions.}
\label{fig:y}
\end{figure}

\begin{figure}[ht]
  \begin{center}
    \setlength{\unitlength}{1cm}
    \begin{picture}(15,10)
      \setlength{\unitlength}{1cm}
      \put(0.5,0){\includegraphics[width=15cm,height=10cm]{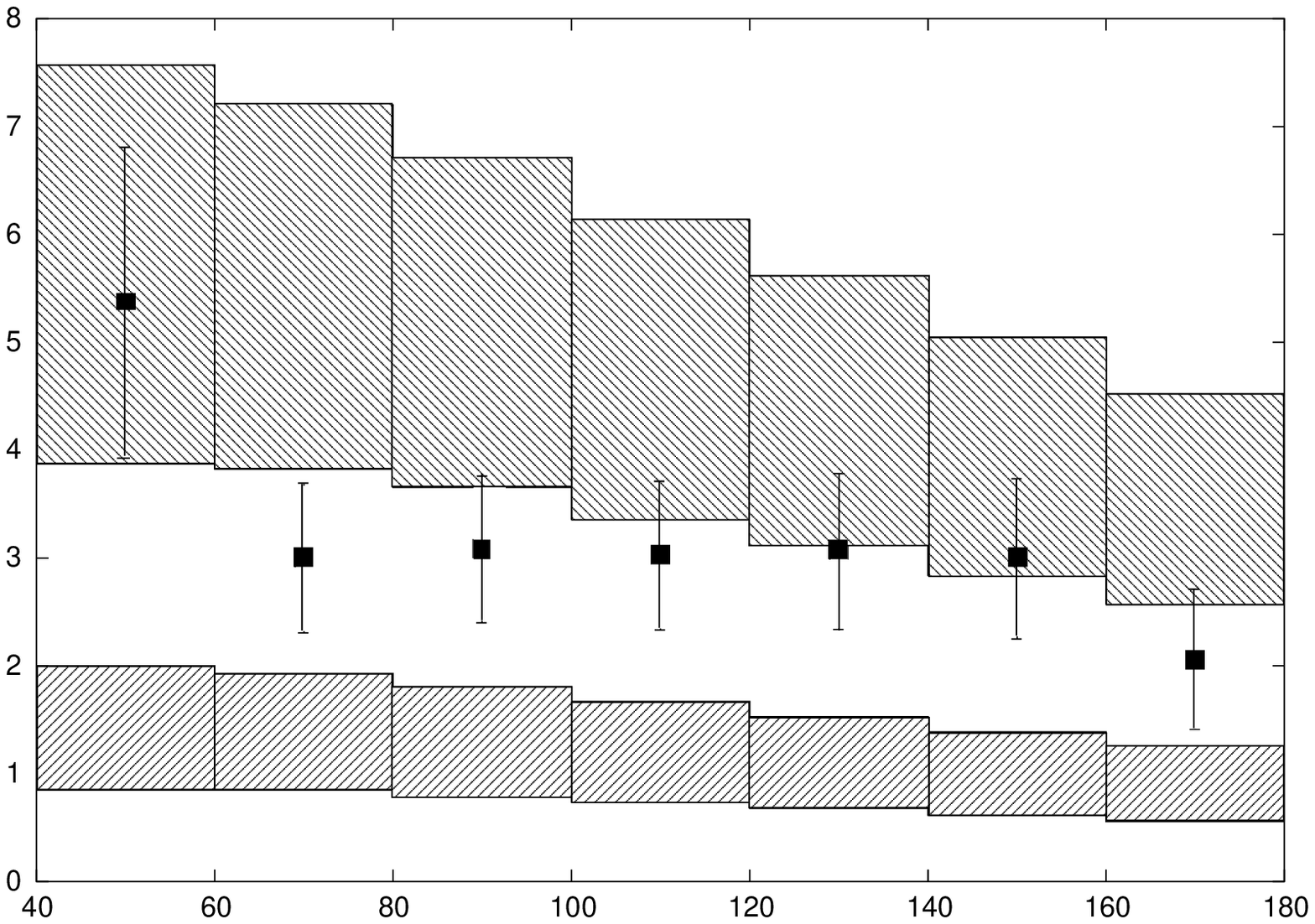}}
      \put(-1.5,5){$\displaystyle\frac{d\sigma}{dW}$}
      \put(-1.5,4){[pb/GeV]}
      \put(7.5,-0.5){$W$ [GeV]}
    \end{picture}
  \end{center}
\caption{The $W$ distribution of $J/\psi$ inclusive production in $ep$ DIS
measured with the H1 detector \protect\cite{mey} is compared with the LO NRQCD
(upper histogram) and CSM (lower histogram) predictions.}
\label{fig:W}
\end{figure}

\begin{figure}[ht]
  \begin{center}
    \setlength{\unitlength}{1cm}
    \begin{picture}(15,10)
      \setlength{\unitlength}{1cm}
      \put(0.5,0){\includegraphics[width=15cm,height=10cm]{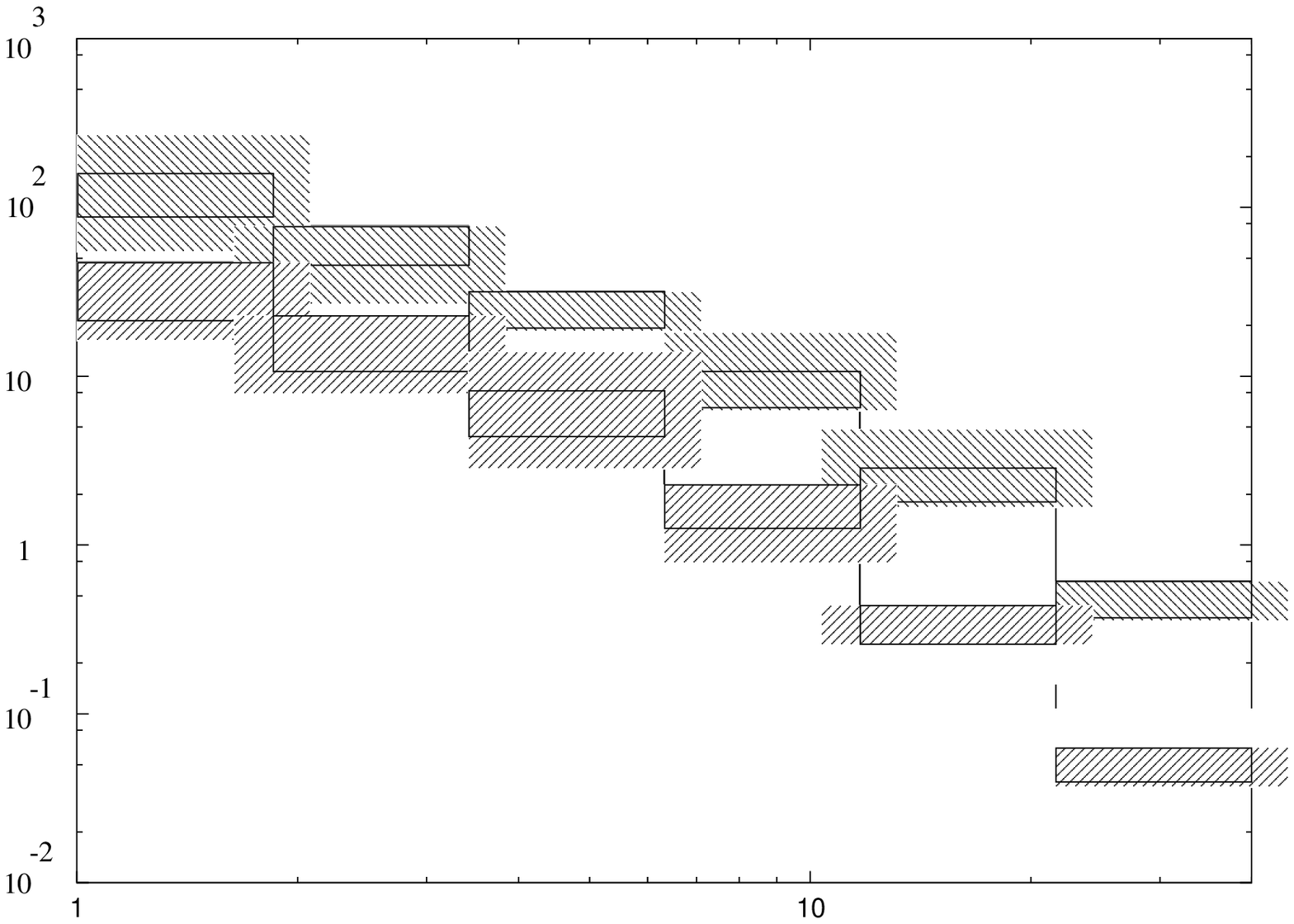}}
      \put(-1.5,5){$\displaystyle\frac{d\sigma}{d\left(p_T^\star\right)^2}$}
      \put(-1.5,4){[pb/GeV${}^2$]}
      \put(7.5,-0.5){$\left(p_T^\star\right)^2$ [GeV${}^2$]}
    \end{picture}
  \end{center}
\caption{LO NRQCD (upper histogram) and CSM (lower histogram) predictions for
the $\left(p_T^\star\right)^2$ distribution of $J/\psi$ inclusive production
in $ep$ DIS.}
\label{fig:pT2}
\end{figure}

\begin{figure}[ht]
  \begin{center}
    \setlength{\unitlength}{1cm}
    \begin{picture}(15,10)
      \setlength{\unitlength}{1cm}
      \put(0.5,0){\includegraphics[width=15cm,height=10cm]{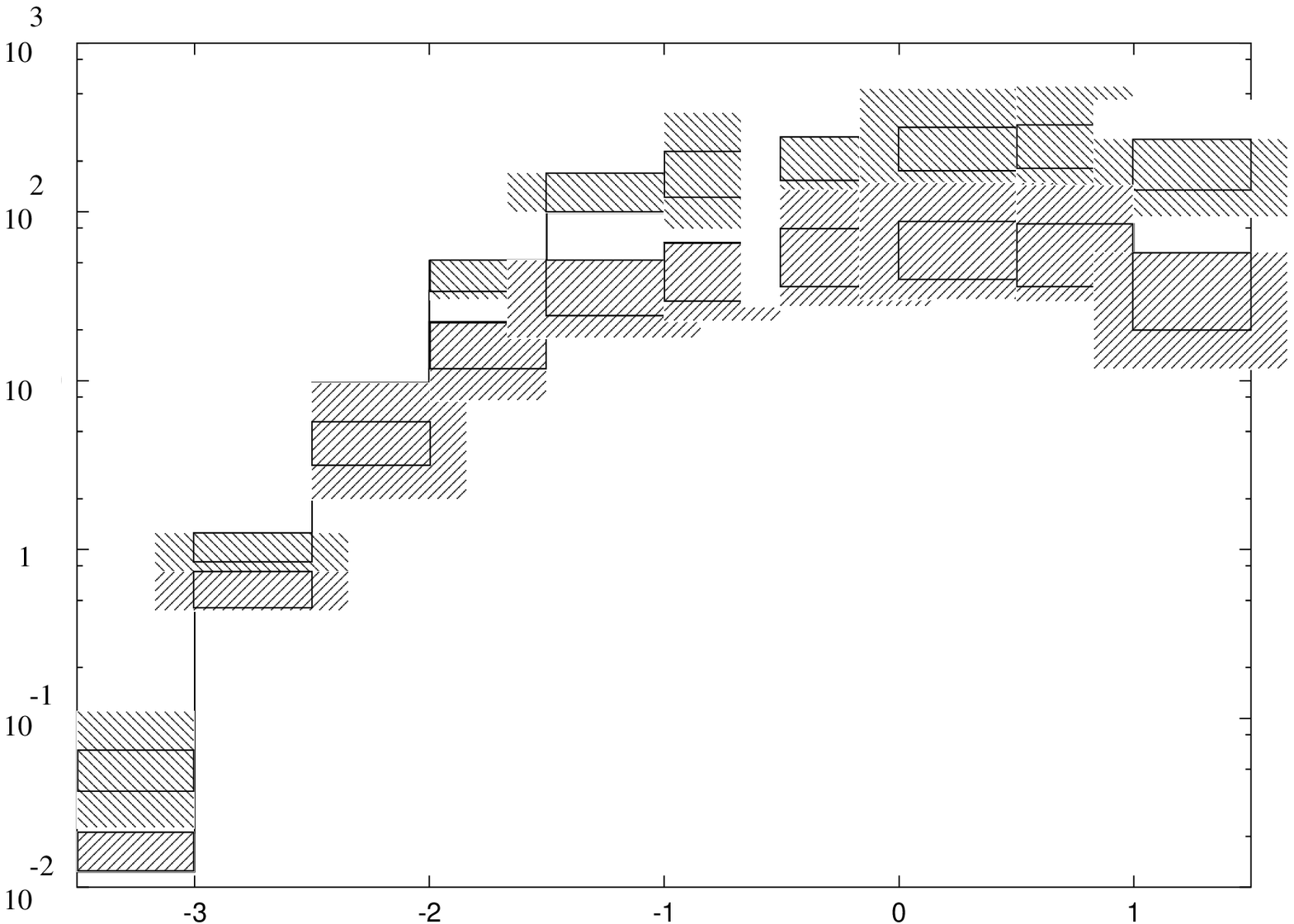}}
      \put(0,5){$\displaystyle\frac{d\sigma}{dy}$}
      \put(0,4){[pb]}
      \put(8.5,-0.5){$y$}
    \end{picture}
  \end{center}
\caption{LO NRQCD (upper histogram) and CSM (lower histogram) predictions for
the $y$ distribution of $J/\psi$ inclusive production in $ep$ DIS.}
\label{fig:yLab}
\end{figure}

In Figs.~\ref{fig:Q2}--\ref{fig:W}, we confront the measured $Q^2$, $p_T^2$,
$z$, $y^\star$, and $W$ distributions, respectively, with our NRQCD
predictions.
For comparison, we also show the corresponding CSM predictions.
In each case, the theoretical errors, evaluated as explained above, are 
indicated by the hatched areas.
Instead of presenting our theoretical predictions as continuous curves, we
adopt the binning pattern encoded in the experimental data, so as to
facilitate quantitative comparisons.
As for the $Q^2$ and $p_T^2$ distributions, the experimental data agrees
rather well with the NRQCD predictions, both in normalization and shape, while
it significantly overshoots the CSM predictions.
In the case of the $z$ distribution, the NRQCD prediction, in general, agrees
better with the experimental data than the CSM one as far as the normalization
is concerned.
As for the shape, however, the experimental measurement favours the CSM
prediction, while the NRQCD one exhibits an excess at large values of $z$,
which is familiar from $J/\psi$ inclusive photoproduction \cite{cac,ko}.
As in the latter case, this rise in $z$ is chiefly due to the
${}^1\!S_0^{(8)}$ and ${}^3\!P_J^{(8)}$ channels.
In the case of the $y^\star$ distribution, the NRQCD prediction nicely agrees
with the experimental data for $y^\star<3$, while it appreciably overshoots
the latter in the very forward direction.
The CSM prediction roughly agrees with the experimental $y^\star$
distribution at the endpoints, while it significantly falls short of the
latter in the central region.
In the case of the $W$ distribution, the experimental data mostly lie in the 
middle between the NRQCD and CSM predictions, slightly favouring the former.
We conclude from Figs.~\ref{fig:Q2}--\ref{fig:W} that the H1 data \cite{mey}
tends to support the NRQCD predictions, while, in general, it overshoots the
CSM predictions.
In Figs.~\ref{fig:pT2} and \ref{fig:yLab}, we present our NRQCD and CSM
predictions for the $\left(p_T^\star\right)^2$ and $y$ distributions,
respectively, although there are no experimental data to compare them with.

At this point, we should compare our numerical results for the cross section
of $e+p\to e+J/\psi+j+X$ with the ones presented in Table~II of
Ref.~\cite{fle}.
To this end, we adopt the theoretical input and kinematic conditions from
Ref.~\cite{fle}.
Specifically, the authors of Ref.~\cite{fle} evaluated $\alpha_s^{(n_f)}(\mu)$
with $n_f=4$ and $\Lambda^{(4)}=130$~MeV, employed the LO proton PDF set by
Gl\"uck, Reya, and Vogt \cite{grv}, took the NRQCD MEs to be
$\left\langle{\cal O}^\psi\left[{}^3\!S_1^{(1)}\right]\right\rangle
=1.1$~GeV${}^3$,
$\left\langle{\cal O}^\psi\left[{}^1\!S_0^{(8)}\right]\right\rangle
=1\times10^{-2}$~GeV${}^3$,
$\left\langle{\cal O}^\psi\left[{}^3\!S_1^{(8)}\right]\right\rangle
=1.12\times10^{-2}$~GeV${}^3$, and
$\left\langle{\cal O}^\psi\left[{}^3\!P_0^{(8)}\right]\right\rangle/m_c^2
=5\times10^{-3}$~GeV${}^3$, and required that $30<W<150$~GeV.
All their other choices, except for the cuts on $Q^2$, $p_T$, $p_T^\star$, and
$z$, coincide with our default settings.
The outcome of this comparison is presented in Table~\ref{tab:com}.
We are unable to determine the source of discrepancy.

\begin{table}[ht]
\begin{center}
\caption{Comparison of our results for the CS and CO contributions to
$\sigma(e+p\to e+J/\psi+j+X)$ in DIS with the ones of Ref.~\protect\cite{fle}.}
\label{tab:com}
\medskip
\begin{tabular}{|c|c|r|r|}
\hline\hline
Type & Cuts & Ref.~\cite{fle} & Our result \\
\hline
CS & $Q^2>4$~GeV${}^2$ & 89~pb & 107~pb \\
CS & $Q^2,p_T^2>4$~GeV${}^2$ & 40~pb & 62~pb \\
CS & $Q^2>4$~GeV${}^2$, $\left(p_T^\star\right)^2>2$~GeV${}^2$, $z<0.8$ &
13~pb & 24~pb \\
CO & $Q^2>4$~GeV${}^2$, $\left(p_T^\star\right)^2>2$~GeV${}^2$, $z<0.8$ &
8~pb & 16~pb \\
\hline\hline
\end{tabular}
\end{center}
\end{table}

Before the advent of the NRQCD factorization formalism \cite{bbl}, one of the
major motivations to study $J/\psi$ inclusive production in $ep$ DIS was to
extract the gluon PDF of the proton $f_{g/p}(x,M)$ \cite{gui,mer}.
In fact, to LO the CSM, the only contributing partonic subprocess is
$e+g\to e+c\overline{c}\left[{}^3\!S_1^{(1)}\right]+g$, and
$\left\langle{\cal O}^\psi\left[{}^3\!S_1^{(1)}\right]\right\rangle$ is well
determined from the partial width of the $J/\psi$ decay to lepton pairs.
Furthermore, $x$ is experimentally accessible through the relation
$x=1-M_{X^\prime}^2/[y(1-z)S]$, and $M=\sqrt{Q^2+M_\psi^2}$ is a plausible
choice.
To LO in NRQCD, this task is somewhat impeded by the presence of CO partonic
subprocesses with quarks or antiquarks in the initial state and by the 
presently still considerable uncertainties in the CO MEs.
We observe from Figs.~\ref{fig:Q2}--\ref{fig:W} that the CO contributions lead
to a dramatic increase in cross section relative to the CSM predictions.
In fact, if we consider the total cross section evaluated with the H1
acceptance cuts \cite{mey}, then the CSM contribution only makes up 25\% of 
the full NRQCD result.
Thus, the present uncertainties in the CO MEs do constitute a serious problem.
However, the quark- and antiquark-induced CO subprocesses only yield a minor
contribution, less than 5\% of the total cross section.
Thus, $J/\psi$ inclusive production in $ep$ DIS remains to be a sensitive
probe of the gluon PDF of the proton if we pass from the CSM to NRQCD.

\section{Conclusions}
\label{sec:four}

We provided, in analytic form, the cross sections of the partonic subprocesses
$e+a\to e+c\overline{c}[n]+a$, where $a=g,q,\overline{q}$ and
$n={}^3\!S_1^{(1)},{}^1\!S_0^{(8)},{}^3\!S_1^{(8)},{}^3\!P_J^{(8)}$, to LO in
the NRQCD factorization formalism \cite{bbl}.
Using these results, we then studied the cross section of
$e+p\to e+J/\psi+j+X$ in DIS under HERA kinematic conditions and compared
various of its distributions with recent H1 data \cite{mey}, from which the
contribution of elastic $J/\psi$ production was separated by a suitable
acceptance cut on $z$.
It turned out that, in general, the experimental data agrees reasonably well 
with our NRQCD predictions, while they tend to disfavour the CSM ones.
However, a familiar feature of $J/\psi$ inclusive photoproduction in $ep$ 
scattering at HERA \cite{cac,ko} carries over to DIS, namely that the cross
section predicted to LO in NRQCD exhibits a distinct rise as $z\to1$, which is
absent in the experimental $z$ distribution.

At this point, it is still premature to jump to conclusions concerning the
experimental verification or falsification of the NRQCD factorization 
hypothesis.
On the one hand, the theoretical predictions for $J/\psi$ inclusive production
in $p\overline{p}$ scattering at the Tevatron \cite{bai,cho} and in $ep$ DIS
at HERA are of LO in $\alpha_s$ and $v$, and they suffer from considerable
uncertainties, mostly from the scale dependences and from the lack of
information on the CO MEs.
On the other hand, the experimental errors are still rather sizeable 
\cite{abe,mey,h1}.
The latter will be dramatically reduced with the upgrades of HERA and the
Tevatron, and with the advent of CERN LHC and hopefully a future $e^+e^-$
linear collider such as DESY TESLA.
On the theoretical side, it is necessary to calculate the NLO corrections to
the hard-scattering cross sections and to include the effective operators
which are suppressed by higher powers of $v$.

%\bigskip
\newpage
\noindent
{\bf Acknowledgements}
\smallskip

\noindent
We acknowledge the collaboration of Jungil Lee at the initial stage of this
work.
We thank Jochen Bartels for instructive comments on the possibilities of
suppressing the contribution from $J/\psi$ diffractive production in $ep$ DIS.
We are grateful to Sean Fleming for helpful communications concerning 
Ref.~\cite{fle}.
We are indepted to Arnd Meyer, Susanne Mohrdieck, and Beate Naroska for
numerous useful discussions regarding Ref.~\cite{mey,h1,mor}.
The work of L.Z. was supported by the Studienstiftung des deutschen Volkes
through a PhD scholarship.
This work was supported in part by the Deutsche Forschungsgemeinschaft through
Grant No.\ KN~365/1-1, by the Bundesministerium f\"ur Bildung und Forschung
through Grant No.\ 05~HT1GUA/4, by the European Commission through the
Research Training Network {\it Quantum Chromodynamics and the Deep Structure
of Elementary Particles} under Contract No.\ ERBFMRX-CT98-0194, and by Sun
Microsystems through Academic Equipment Grant No.~EDUD-7832-000332-GER.

\def\theequation{\Alph{section}.\arabic{equation}}
\begin{appendix}
\setcounter{equation}{0}
\section{Partonic cross sections}

In this Appendix, we present analytic expressions for the coefficients
$F_a[n]$, $T_a[n]$, and $L_a[n]$ appearing in Eq.~(\ref{eq:res}).
In order to compactify the expressions, it is useful to introduce the Lorentz
invariants $s=2q\cdot p$, $t=-2p\cdot p^\prime$, and $u=-2q\cdot p^\prime$,
which are related to the partonic Mandelstam variables by $s=\hat s+Q^2$,
$t=\hat t$, and $u=\hat u+Q^2$, respectively.
In the following, $e_q$ is the fractional electric charge of quark $q$.

$e+q(\overline{q})\to c\overline{c}\left[{}^3\!S_1^{(1)}\right]
+q(\overline{q})$:
\begin{equation}
F=T=L=0.
\end{equation}

$e+q(\overline{q})\to c\overline{c}\left[{}^1\!S_0^{(8)}\right]
+q(\overline{q})$:
\begin{eqnarray}
F&=&\frac{-16 \pi^2 e_c^2 \alpha \alpha_s^2}{3 M_\psi s^4 t (s + u)^2},
\nonumber\\
T&=&2 Q^4 t^2+2 Q^2 s t (s + u)+s^2 (s^2 + u^2),
\nonumber\\
L&=&-2 t (Q^2 t + s u).
\end{eqnarray}

$e+q(\overline{q})\to c\overline{c}\left[{}^3\!S_1^{(8)}\right]
+q(\overline{q})$:
\begin{eqnarray}
F&=&\frac{-8 \pi^2 e_q^2 \alpha \alpha_s^2}{9 M_\psi^3 s^4 (Q^2 - s)^2  
  (Q^2 - u)^2},
\nonumber\\
T&=&2 Q^6 t^2 (2 s + t)+
    2 Q^4 s [s^2 u - s t (3 t - 2 u) - 2 t^3]+
    Q^2 s^2 [s^2 (t - 2 u)
\nonumber\\
&&{} + 2 s (t^2 - 4 t u - u^2) + t (2 t^2 - 2 t u - u^2)]+
    s^3 u (s^2 + 2 s t + 2 t^2 + 2 t u + u^2),
\nonumber\\
L&=&-2 (Q^2 - s)^2 t (s + t)^2.
\end{eqnarray}

$e+q(\overline{q})\to c\overline{c}\left[{}^3\!P_J^{(8)}\right]
+q(\overline{q})$:
\begin{eqnarray}
F&=&\frac{64 \pi^2 e_c^2 \alpha \alpha_s^2}{3 M_\psi^3 s^4 t (s + u)^4},
\nonumber\\
T&=&8 Q^6 t [2 s^2 + s t + t (2 t + u)]-
    2 Q^4 [4 s^4 + 12 s^3 t + s^2 (19 t^2 + 8 t u + 4 u^2)
    + 2 s t (6 t^2 
\nonumber\\
&&{} + t u - 2 u^2) + t^2 (8 t^2 + 12 t u + 7 u^2)]+
    2 Q^2 s [6 s^4 + s^3 (7 t + 6 u)+ 
    s^2 (4 t^2 + 3 t u
\nonumber\\
&&{}+ 6 u^2) - s u (8 t^2 + 3 t u - 6 u^2)  
    - t u (8 t^2 + 12 t u + 7 u^2)]-s^2 (s + u) 
    [7 s^3 + s^2 (12 t
\nonumber\\
&&{}+ 7 u) + 
    s (8 t^2 + 16 t u + 7 u^2) + u (8 t^2 + 12 t u + 7 u^2)],
\nonumber\\
L&=&8 Q^4 t [s^2 - s t - t (2 t + u)]-
    2 Q^2 [2 s^4 + 4 s^3 t + s^2 (5 t^2 + 8 t u + 2 u^2)
    - 2 s t (6 t^2
\nonumber\\
&&{}+ t u - 2 u^2) - t^2 (8 t^2 + 12 t u + 7 u^2)]+
    2 s (s + u) [2 s^3 + 2 s^2 t + s (8 t^2 + 5 t u
\nonumber\\
&&{} + 2 u^2) + t (8 t^2 + 12 t u + 7 u^2)].
\end{eqnarray}

$e+g\to c\overline{c}\left[{}^3\!S_1^{(1)}\right]+g$:
\begin{eqnarray}
F&=&\frac{64 \pi^2 e_c^2 \alpha \alpha_s^2}{27 M_\psi s^4 (s + t)^2 
    (s + u)^2 (t + u)^2},
\nonumber\\
T&=&-4 Q^6 t^2 (s^2 + t^2)+
    2Q^4 t [s^3 (3 t - 2 u) + 3 s^2 t (t + u) + 2 s t^2 (t - u) 
    + 2 t^3 (t + u)]
\nonumber\\
&&{}-2Q^2 s [s^3 (t - u)^2 - 2 s^2 t u (t + u) - s t^2 u (2 t - u)
    - 2 t^3 u (t + u)] +
    2s^2 [ s^3 (t^2
\nonumber\\
&&{} + t u + u^2) + s^2 (t + u)^3 + 
    s t u (t^2 + 3 t u + u^2) + t^2 u^2 (t + u)],
\nonumber\\
L&=& -2Q^4 t^2 (s^2 - 2 t^2)+ 2Q^2 t (s^2 - 2 t^2) [s (t - u) 
    + t (t + u)]
    - s [s^3 (t^2 + u^2)
\nonumber\\
&&{} + 2 s^2 t^2 (t + u) 
      + s t^2 (t^2 + 6 t u + u^2)+ 4 t^3 u (t + u)].
\end{eqnarray}

$e+g\to c\overline{c}\left[{}^1\!S_0^{(8)}\right]+g$:
\begin{eqnarray}
F&=&\frac{24 \pi^2 e_c^2 \alpha \alpha_s^2}{M_\psi s^4 t (s + t)^2 
    (s + u)^2 (t + u)^2},
\nonumber\\
T&=& 2 Q^6 t^3 u^2 + 2 Q^4 s t^2 u [s (t + u) + t^2 + t u + 2 u^2
     ] + 
     Q^2 s^2 t [s^4 + 2 s^3 (t + u)
\nonumber\\
&&{} + 3 s^2 (t + u)^2
     + 2 s (t^3 + 3 t^2 u + 4 t u^2 + 2 u^3) + t^4 + 2 t^3 u + 
     5 t^2 u^2 + 4 t u^3 + 3 u^4]
\nonumber\\
&&{} + s^3 u [s^4 + 2 s^3 (t + u)
     + 3 s^2 (t + u)^2 + 2 s (t + u)^3 + (t^2 + t u + u^2)^2],
\nonumber\\
L&=& -2 Q^4 t^3 u^2 -
     2 Q^2 s t^2 u [ s (t + u) + t^2 + t u + 2 u^2]-
      s^2 t [s^2 (t + u)^2
\nonumber\\
&&{} + 2 s (t^3 + 2 t^2 u + 2 t u^2 + u^3)
     + t^4 + 2 t^3 u + 3 t^2 u^2 + 2 t u^3 + 2 u^4].
\end{eqnarray}

$e+g\to c\overline{c}[{}^3\!S_1^{(8)}]+g$:
\begin{eqnarray}
F&=&\frac{15}{8}
F\left(e+g\to c\overline{c}\left[{}^3\!S_1^{(1)}\right]+g\right),
\nonumber\\
T&=&T\left(e+g\to c\overline{c}\left[{}^3\!S_1^{(1)}\right]+g\right),
\nonumber\\
L&=&L\left(e+g\to c\overline{c}\left[{}^3\!S_1^{(1)}\right]+g\right).
\end{eqnarray}

$e+g\to c\overline{c}\left[{}^3\!P_J^{(8)}\right]+g$:
\begin{eqnarray}
F&=&\frac{96 \pi^2 e_c^2 \alpha \alpha_s^2}{M_\psi^3 s^4 t (s + t)^3 
   (s + u)^4 (t + u)^3},
\nonumber\\
T&=& -8 Q^8 t^2 [2 s^5 t - 2 s^4 u^2 + s^3 t^2 (2 t - 3 u) - s^2 t u (t^2 
     + 3 t u - 3 u^2) - s t u (2 t^3 + t^2 u - t u^2
\nonumber\\
&&{} - u^3) + t^2 u^3 (2 t + u)] +
     2 Q^6 t [4 s^7 (t + u) + 4 s^6 (5 t^2 + 2 u^2) + 
     4 s^5 (5 t^3 + 6 t^2 u - 2 t u^2
\nonumber\\
&&{} + 4 u^3) 
     + 2 s^4 (9 t^4 - 9 t^3 u + 4 t^2 u^2 - 6 t u^3 + 4 u^4)
     + s^3 (12 t^5 - 2 t^4 u - 29 t^3 u^2 + 33 t^2 u^3
\nonumber\\
&&{} - 12 t u^4 + 4 u^5)
     - s^2 t (2 t^5 + 20 t^4 u + 25 t^3 u^2 - t^2 u^3 - 16 t u^4 + 4 u^5)
     - s t^2 u (12 t^4 + 26 t^3 u
\nonumber\\
&&{} + 2 t^2 u^2 + t u^3 + u^4)
     - t^3 u^2 (2 t^3 - 6 t^2 u - 15 t u^2 - 7 u^3)]-
     2 Q^4 [2 s^8 (3 t^2 + t u - 2 u^2)
\nonumber\\
&&{} + s^7 (21 t^3 + 3 t^2 u + 10 t u^2 - 8 u^3)
     + s^6 (28 t^4 + 15 t^3 u - 15 t^2 u^2 + 22 t u^3 - 12 u^4)
     + s^5 (27 t^5
\nonumber\\
&&{} + 11 t^4 u + 8 t^3 u^2 - 20 t^2 u^3 + 26 t u^4 - 8 u^5)
     + s^4 (16 t^6 + 17 t^5 u - 20 t^4 u^2 + 2 t^3 u^3 - 33 t^2 u^4
\nonumber\\ 
&&{} + 18 t u^5 - 4 u^6)
     + s^3 t (2 t^6 + 8 t^5 u - 3 t^4 u^2 - 19 t^3 u^3 - 19 t^2 u^4 - 17 t u^5 
     + 6 u^6)    
     - s^2 t^2 (2 t^6
\nonumber\\
&&{} + 14 t^5 u + 30 t^4 u^2 + 33 t^3 u^3 + 
     46 t^2 u^4 + 21 t u^5 + 4 u^6)    
     - 2 s t^3 u (2 t^5 + 10 t^4 u + 16 t^3 u^2
\nonumber\\
&&{} + 20 t^2 u^3 + 19 t u^4 + 7 u^5)    
     - 2 t^6 u^2 (t + u)^2] +
     Q^2 s [ s^8 (7 t^2 - 5 t u - 12 u^2)
     + s^7 (25 t^3 + 3 t^2 u
\nonumber\\
&&{} - 18 t u^2 - 36 u^3)
     + s^6 (37 t^4 + 25 t^3 u - 12 t^2 u^2 - 20 t u^3 - 60 u^4)
     + s^5 (39 t^5 + 39 t^4 u + 16 t^3 u^2
\nonumber\\
&&{} - 18 t^2 u^3 - 6 t u^4 - 60 u^5)
     + s^4 (29 t^6 + 83 t^5 u + 72 t^4 u^2 + 88 t^3 u^3 + 40 t^2 u^4 + 
     22 t u^5 - 36 u^6)
\nonumber\\
&&{} + s^3 (9 t^7 + 75 t^6 u + 148 t^5 u^2 + 176 t^4 u^3 + 178 t^3 u^4 + 
     102 t^2 u^5 + 22 t u^6 - 12 u^7)
     + s^2 t u (22 t^6
\nonumber\\
&&{} + 107 t^5 u + 199 t^4 u^2 + 
     211 t^3 u^3 + 177 t^2 u^4 + 73 t u^5 + 9 u^6)
     + s t^2 u^2 (17 t^5 + 69 t^4 u + 107 t^3 u^2
\nonumber\\
&&{} + 105 t^2 u^3 + 71 t u^4 + 
     21 u^5)    
     + 4 t^5 u^3 (t + u)^2] + 
     s^2 (s + u) [7 s^7 u (t + u) 
     + s^6 u (25 t^2 + 38 t u
\nonumber\\
&&{} + 21 u^2)
     + s^5 (2 t^4 + 47 t^3 u + 88 t^2 u^2 + 78 t u^3 + 35 u^4)
     + s^4 (4 t^5 + 63 t^4 u + 132 t^3 u^2
\nonumber\\
&&{} + 156 t^2 u^3 + 98 t u^4 + 35 u^5)
     + s^3 (2 t^6 + 47 t^5 u + 136 t^4 u^2 + 190 t^3 u^3 + 156 t^2 u^4 
     + 78 t u^5
\nonumber\\ 
&&{} + 21 u^6)
     + s^2 u (13 t^6 + 70 t^5 u + 136 t^4 u^2 + 132 t^3 u^3 + 88 t^2 u^4 + 
     38 t u^5 + 7 u^6)
     + s t u^2 (13 t^5
\nonumber\\
&&{} + 47 t^4 u + 63 t^3 u^2 + 47 t^2 u^3 + 25 t u^4 + 7 u^5)
     + 2 t^4 u^3 (t + u)^2],
\nonumber\\
L&=& -8 Q^6 t^2 [s^5 t- s^4 u^2- 2 s^3 t^3 + s^2 t^2 u (t 
     + 3 u) 
     + s t u (2 t^3 + t^2 u - t u^2 - u^3)
     - t^2 u^3 (2 t + u)]
\nonumber\\
&&{} + 2 Q^4 t [2 s^7 (t + u) + 4 s^6 (2 t^2 + u^2) 
     - s^5 (t^3 - t^2 u + 2 t u^2 - 8 u^3)
     - s^4 (21 t^4 + 13 t^3 u
\nonumber\\
&&{} + 18 t^2 u^2 - 2 t u^3 - 4 u^4)
     - 2 s^3 (6 t^5 + 10 t^4 u + t^3 u^2 + 4 t^2 u^3 - 6 t u^4 - u^5)
     + 2 s^2 t (t^5 + 10 t^4 u
\nonumber\\
&&{} + 11 t^3 u^2 + 9 t^2 u^3 + 6 t u^4 + 5 u^5)
     + s t^2 u (12 t^4 + 26 t^3 u + 2 t^2 u^2 + t u^3 + u^4)
     + t^3 u^2 (2 t^3
\nonumber\\
&&{} - 6 t^2 u - 15 t u^2 - 7 u^3)] - 
     4 Q^2 [s^8 (t^2 - u^2)
     + s^7 (3 t^3 + t u^2 - 2 u^3)
     - s^6 (2 t^4 + 2 t^3 u + 2 t^2 u^2
\nonumber\\
&&{} - t u^3 + 3 u^4)
     - s^5 (12 t^5 + 16 t^4 u + 16 t^3 u^2 - t^2 u^3 + t u^4 + 2 u^5)
     - s^4 (10 t^6 + 28 t^5 u + 28 t^4 u^2
\nonumber\\
&&{} + 13 t^3 u^3 - 7 t^2 u^4 + 3 t u^5 + 
     u^6)
     - s^3 t (t^6 + 9 t^5 u + 18 t^4 u^2 + t^3 u^3 - 15 t^2 u^4 - 
     10 t u^5 + 2 u^6)
\nonumber\\
&&{} + s^2 t^2 (t^6 + 7 t^5 u + 13 t^4 u^2 + 18 t^3 u^3 + 
     35 t^2 u^4 + 24 t u^5 + 6 u^6)    
     + s t^3 u (2 t^5 + 10 t^4 u + 16 t^3 u^2
\nonumber\\
&&{} + 20 t^2 u^3 + 19 t u^4 + 7 u^5)    
     + t^6 u^2 (t + u)^2] -
     2 s (s + u) [2 s^7 u (t + u)
     + 2 s^6 u (3 t^2 + 3 t u
\nonumber\\
&&{} + 2 u^2) 
     + s^5 (5 t^4 + 15 t^3 u + 18 t^2 u^2 + 14 t u^3 + 6 u^4)
     + s^4 (15 t^5 + 38 t^4 u + 53 t^3 u^2 + 40 t^2 u^3
\nonumber\\
&&{} + 22 t u^4 + 4 u^5)
     + s^3 (15 t^6 + 52 t^5 u + 88 t^4 u^2 + 81 t^3 u^3 + 47 t^2 u^4 
     + 19 t u^5 + 2 u^6)+ s^2 t (5 t^6
\nonumber\\
&&{} + 32 t^5 u + 78 t^4 u^2 + 90 t^3 u^3 + 68 t^2 u^4 + 34 t u^5 
     + 9 u^6)
     + s t^2 u (7 t^5 + 31 t^4 u + 47 t^3 u^2 + 39 t^2 u^3
\nonumber\\
&&{} + 23 t u^4 + 7 u^5)
     + 2 t^5 u^2 (t + u)^2].
\end{eqnarray}

\end{appendix}

\newpage

\newpage

\newpage

\newpage

\newpage

\newpage

\newpage

\newpage

\newpage

\end{document}